\documentclass[11pt]{article}

\usepackage{amsmath}
\usepackage{amsfonts}
\usepackage{amssymb}
\usepackage{graphicx}
\usepackage{epstopdf}
\usepackage{cite}
\usepackage[usenames]{color}

\providecommand{\U}[1]{\protect\rule{.1in}{.1in}}
\newcommand{\ba}{\begin{array}}
\newcommand{\ea}{\end{array}}
\newcommand{\Dsl}[1] { \setbox0=\hbox{$#1$}     
\dimen0=\wd0   \setbox1=\hbox{/} \dimen1=\wd1  \ifdim\dimen0>\dimen1        
 \rlap{\hbox to \dimen0{\hfil/\hfil}}  #1 \else \rlap{\hbox to \dimen1{\hfil$#1$\hfil}}  /  \fi  }
\newcommand{\bea}{\begin{eqnarray}}
\newcommand{\eea}{\end{eqnarray}}

\newcommand{\ns}{\Dsl{n}}
\newcommand {\nbs}{\Dsl{\bar n}}
\newcommand{\nbn}{\frac{\nbs\ns}{4}}
\newcommand{\nnb}{\frac{\ns\nbs}{4}}

\newcommand{\lab}{\label}
\newcommand{\NEG}[1]{\Dsl{#1}} 

\begin{document}

\title{ {\Large \bf A Study of  Power Suppressed  Contributions  in  $J/\psi\rightarrow p\bar{p}$ Decay }}
\author{ Nikolay Kivel 
\\[1cm]
\textit{Petersburg Nuclear Physics Institute, 188300, Gatchina, Russia} }

\maketitle

\begin{abstract}
The power suppressed amplitude which describes the Pauli ($\sigma^{\mu\nu}$) coupling  in the $J/\psi\to p\bar p$ decay  
is calculated  within the effective field theory framework.  It is shown that at the leading-order approximation  this contribution  
is factorisable and the overlap with the hadronic final state can be described by collinear matrix elements.  
The obtained contribution depends on the nucleon light-cone distribution amplitudes of twist-3 and twist-4.   
This result is used for a qualitative phenomenological analysis of existing data for  $J/\psi\to p\bar p$ decay:  branching ratio and 
 the angular distribution in the cross section $e^+e^-\to J/\psi\to p\bar p$.   
\end{abstract}

\noindent

\vspace*{1cm}

\newpage

\section{Introduction}

\label{int}

Hadronic exclusive decays of heavy charmonia remain one of the most challenging
subject for theoretical investigations, see for instance reviews in Refs. \cite{Brambilla:2004wf, Brambilla:2010cs}. 
The effective field theory framework  provides the most effective approach for
calculations of the corresponding decay amplitudes. Such a framework
involves the nonrelativistic expansion with respect to a small heavy quark
velocity $v$ \cite{Bodwin:1994jh, Lepage:1992tx}  and the expansion  with respect to a small ratio $\lambda
^{2}\sim \Lambda /m_{Q}$ where $\Lambda $ is the typical hadronic scale.  A
decay amplitude  can be represented as a superposition of the  hard  and
soft contributions  associated with the  hard and soft scales,
respectively. The nonperturbative long distance dynamics is described  by the well defined, process independent matrix elements. 
The nice feature of the  effective field theory framework is the systematic power  of the various counting contributions  
with respect to small parameters $v$ and $\lambda $. This allows one to perform a
systematic calculations and better understand complicated underlying partonic dynamics. 

The power counting is closely associated  with the hard  partonic subprocess and helicities of the initial and final
hadrons.  This  leads to the well known helicity selection rule, see {\it e.g.} Refs. \cite{Brodsky:1981kj, Chernyak:1983ej}.
 A typical feature of hard processes  is  that their amplitudes  are dominated by contributions with 
 the helicity conserving partonic amplitudes. The partonic configurations, which require a helicity flip or involve the  angular momentum, 
  are  suppressed by the powers of the small $\lambda$.  Nevertheless,  there are many indications   that corresponding subleading 
  amplitudes might be relevant for an understanding  of  many exclusive  decays.
 In particular such power suppressed corrections  are required for a description of  charmonium  
 decays into  baryon-antibaryon pair \cite{Brambilla:2004wf}.

The first calculation of such charmonia decay  was done in Ref.\cite{Brodsky:1981kj}.  
Later the various decays of $S$- and $P$- wave charmonia were  studied in many publications, 
see Refs. \cite{Brambilla:2004wf, Brambilla:2010cs, Chernyak:1983ej} and references therein.
  In all of these calculations  the  subleading  amplitudes as a rule  have been neglected. 
  However,  in some  cases such a naive estimate does not agree with  experimental data. 
  For instance, the ratio $\Gamma \lbrack \chi _{c0}\rightarrow p\bar{p}]/\Gamma \lbrack \chi _{c1}\rightarrow p\bar{p}
]\simeq 36$   \cite{Tanabashi:2018oca}  is very large despite the  $\chi _{c0}$ 
 amplitude  is  suppressed by small $\lambda ^{2}$.  An understanding  of such effects definitely
requires a careful  study of subleading power corrections.

In this paper we consider the effect of  power suppressed corrections in the description of  $J/\psi \rightarrow p\bar{p}$ decay. 
Such a reaction is simpler than the mentioned  $P$-wave decays  because  $1S$- charmonium state is  a ground  state.
As in the case of the electromagnetic source  the corresponding decay  amplitude is described by the Dirac ($\gamma^\mu$) and by the Pauli ($\sigma^{\mu\nu}$) vertices  and   the second one is suppressed according to pQCD helicity selection rule.  Therefore in many  theoretical considerations  this amplitude is often discarded.

On the other hand,  this amplitude can provide a substantial effect  in  description  of the  polar angular distribution of baryon-antibaryon pairs 
produced in the exclusive decay of the $J/\psi$.   The corresponding angular distribution can be
written as a function of the angle $\theta $ between the nucleon or
antinucleon direction and the beam as follows:
\begin{equation}
\frac{dN}{d\cos \theta }=\mathcal{N}(1+\alpha \cos \theta ),
\label{alpha:def}
\end{equation}%
where $\mathcal{N}$ is an overall normalisation.  The coefficient $\alpha
=1 $ in the limit of infinite mass of heavy quark \cite{Brodsky:1981kj}.  The simple
kinematical effect from the nucleon mass  $m_{N}/M_{\psi }\neq 0$  yields
\cite{Claudson:1981fj}%
\begin{equation}
\alpha =\frac{1-4m_{N}^{2}/M_{\psi }^{2}}{1+4m_{N}^{2}/M_{\psi }^{2}}\simeq
0.455.  \label{alfkin}
\end{equation}%
The angular distribution  was measured in many experiments \cite{Peruzzi:1977pb, Brandelik:1979hy, Eaton:1983kb, Pallin:1987py, Bai:2004jg, Ambrogiani:2004uj, Ablikim:2006aw, Ablikim:2012eu}. The most accurate
measurements \cite{Ablikim:2012eu} give  the value $\alpha \simeq 0.59\pm0.01$. The
difference with the simple prediction in Eq.(\ref{alfkin})  can only be explained by  the  amplitude associated with the
Pauli coupling ($\sigma ^{\mu \nu }$).   The more sophisticated phenomenological  models with massive
constituent quarks were considered in Refs.\cite{Carimalo:1985mw, Murgia:1994dh, Ping:2002uj}.  Despite  various
 assumptions about the QCD underlying dynamics, these calculations gives a reliable description of the  angular distribution. 
A  calculation of  the angular coefficient $\alpha$ within the systematic framework does not depend on a model  of hadron dynamics   
and  might help better understand the role of the power suppressed contributions. 

In the present work the  subleading amplitude is computed
within the effective field theory framework.  Such calculation involves the twist-4  light-cone distributions amplitudes of the nucleon 
which can be associated with the three-quark component of the nucleon wave function. The required
nucleon  matrix elements have been already studied using QCD sum rules in Refs. \cite{Braun:2000kw, Braun:2006hz, Anikin:2013aka} 
and in the lattice calculations \cite{Bali:2015ykx,Bali:2019ecy}. The
interesting observation which can be done from these results is that  the
twist-4 matrix elements associated with the three quark in $P$-wave
configuration are quite large comparing to the leading twist-3 matrix
element. This can lead to a large power corrections because  charmonium mass is not large enough.

The paper is organised as follows. In Sec.\ref{kin} we introduce notations,
kinematics and  provide the known leading-twist results for the decay
amplitude. In Sec.\ref{calc}  we briefly describe the calculation of the
subleading amplitude $A_{2}$ and provide the corresponding analytical results.  We show that
for the $S$-wave charmonia such amplitude is also factorisable and is described
by the well defined convolution integral of the hard partonic amplitude with
the light-cone distribution amplitudes (LCDAs) of twist-3 and twist-4.  The
obtained result is used with various models of the LCDAs  for the qualitative numerical estimates  in Sec.\ref{phen}.  
In Sec.\ref{discuss} we discuss the obtained results.  In
Appendices \ref{AppA}  and \ref{AppB} we provide the information about the nonperturbative matrix elements  and 
 discuss useful technical details.

\section{Definitions, kinematics and the leading-twist amplitude}

\label{kin}

It is convenient to describe the decay $J/\psi (P)\rightarrow p(k)\bar{p}%
(k^{\prime })$ in the charmonia rest frame 
\begin{equation}
P=M_{\psi }\omega ,~\omega =(1,\vec{0}).\ 
\end{equation}%
The outgoing  momenta $k$ and $k^{\prime }$ are directed along the $z$%
-axis and read%
\begin{equation}
k=(M_{\psi }/2,0,0,M_{\psi }\beta/2 ),~ 
k^{\prime }=(M_{\psi }/2,0,0,-M_{\psi }\beta/2 ),~ \beta =\sqrt{1-\frac{4m_{N}^{2}}{M_{\psi }^{2}}},  \label{beta}
\end{equation}%
where $m_{N}$ is the  nucleon mass.  We use  the auxiliary light-cone vectors 
\begin{equation}
n=(1,0,0,-1),~\ \bar{n}=(1,0,0,1).
\end{equation}%
Any four-vector $V$ can be expanded as 
\begin{equation}
V=V_{+}\frac{\bar{n}}{2}+V_{-}\frac{n}{2}+V_{\bot },~
\end{equation}%
where $\ V_{+}=(Vn)=V_{0}+V_{3}$, $V_{-}=(V\bar{n})=V_{0}-V_{3}$.  The
light cone expansions of particle momenta are given by%
\begin{equation}
P=M_{\psi }\frac{1}{2}(n+\bar{n}),~\ k\simeq M_{\psi }\frac{\bar{n}}{2},~
k^{\prime }\simeq M_{\psi }\frac{n}{2}.
\end{equation}

The decay amplitude $J/\psi \rightarrow p\bar{p}$ is defined as%
\begin{equation}
\left\langle k,k^{\prime }\right\vert ~i\hat{T}~\left\vert P\right\rangle
=i(2\pi )^{4}\delta (P-k-k^{\prime })~M,
\end{equation}%
wit 
\begin{equation}
M=\bar{N}(k)\left\{ A_{1}~\Dsl{\epsilon}_{\psi }+A_{2}\left( \epsilon _{\psi
}\right) _{\mu }(k^{\prime }+k)_{\nu }\frac{i\sigma ^{\mu \nu }}{2m_{N}}%
\right\} V(k^{\prime })~,  \label{def:M}
\end{equation}%
where $\Dsl p=p_\mu \gamma^\mu$.
The nucleon $\bar{N}(k)$ and antinucleon $V(k^{\prime })$ spinors have
standard normalisation $\bar N N=2m_N$ and $\bar V V=-2m_N$.  The charmonium polarisation vector $\epsilon _{\psi }^{\mu }\equiv \epsilon
_{\psi }^{\mu }(P,\lambda )$  satisfies 
\begin{equation}
\sum_{\lambda }\epsilon _{\psi }^{\mu }(P,\lambda )\epsilon _{\psi }^{\nu
}(P,\lambda )=-g^{\mu \nu }+\frac{P^{\mu }P^{\nu }}{M_{\psi }^{2}}.
\end{equation}
The scalar amplitudes $A_{1}$ and $A_{2}$  describe the decay process. 
 Within the effective field theory,
where the mass of the heavy quark is much large then the typical hadronic
scale $m_{Q}\gg \Lambda $,  these amplitudes can be computed expanding over
the small relative heavy quark velocity $v$ and over the small ratio $\lambda ^{2}\sim \Lambda /m_{Q}$.  The power counting predicts that the
amplitude $A_{2}$ is suppressed  as
\begin{equation}
A_{2}/A_{1}\sim \lambda ^{2}.
\end{equation}%
Therefore this amplitude is usually neglected. The leading-order in $\alpha_s$ expression for the amplitude $A_{1}$ was calculated a long time ago in Refs. \cite{Brodsky:1981kj,Chernyak:1984bm, Chernyak:1987nv} 
and can be written as \cite{Chernyak:1987nv}
\begin{equation}
A_{1}^{(0)}=\frac{f_{\psi }}{~m_{Q}^{2}}\frac{~f_{N}^{~2}}{m_{Q}^{4}}~(\pi
\alpha _{s})^{3}~\frac{10}{81}~I_{0},  \label{A1:res}
\end{equation}%
where  the collinear convolution integral  is given by
\begin{eqnarray}
I_{0} =\frac{1}{4}\int Dy_{i}\frac{1}{y_{1}y_{2}y_{3}} \int Dx_{i}\frac{1}{x_{1}x_{2}x_{3}}
\left\{ \frac{y_{1}x_{3}}
{ D_1 D_3 }\varphi_{3}(y_{i})~\varphi _{3}(x_{i})
 +\frac{2y_{1}x_{2}}{D_1 D_2 } T_{1}(y_{i})T_{1}(x_{i}) \right\} .
 \label{def:I} 
\end{eqnarray}
with  $D_i=x_i+y_i-2x_i y_i$ and $\varphi _{3}(x_{i})\equiv \varphi _{3}(x_{1},x_2,x_3)$.
The  couplings $f_{\psi }$ and $f_{N}$ in Eq.(\ref{A1:res}) are related
with the long distance matrix elements of charmonia and  nucleon,
respectively. The explicit definitions are given in Appendix A. The nucleon
light-cone distribution amplitudes  $\varphi _{3}(y_{i})$ and $T_{1}(x_{i})$ are
related with the three quark component of the nucleon wave function and
describe the distribution of the quark momenta in the nucleon wave function
at zero transverse separation. They depend on the  quark light-cone
fractions $0< x_{i}<1$ which satisfy  momentum conservation condition
$x_{1}+x_{2}+x_{3}=1$.
Therefore the convolution integrals in Eq.(\ref{A1:res}) have a $\delta $%
-function in the measure%
\begin{equation}
Dx_{i}=dx_{1}dx_{2}dx_{3}\delta (1-x_{1}-x_{2}-x_{3}).
\end{equation}%
The LCDA $T_1$ in Eq.(\ref{def:I}) is  not independent and  related with  $\varphi _{3}$  by Eq.(\ref{T1[phi3]}).
Hence the amplitude $A_{1}$ depends on the one twist-3 LCDA $\varphi _{3}$.  The different
models for this nonperturbative function  will be discussed later.  

From  Eq.(\ref{A1:res}) it follows that $A_{1}\sim v^{3}\lambda ^{8}\alpha
_{s}^{3}$ which is obtained from the fact that $f_{N}\sim \Lambda ^{2}$ and
in NRQCD $f_{\psi }\sim m_{Q}^{2}~v^{3}$. The scaling behaviour of order $v^{3}$ 
is the minimal possible power behaviour according to  NRQCD power scaling.
The helicity flip amplitude $A_{2}$ is suppressed by additional power  $\lambda^{2}$ 
due to the twist-4 nucleon LCDA which can be associated with the three
quarks in $P$-wave state or with the 3-quark-gluon component of the wave
function. Only such configurations can provide a description of the long
distance collinear overlap with the nucleon state in this case. This is a
direct consequence of the helicity conservation in the hard subprocess. In
many hard processes such subleading amplitudes often do not possess collinear
factorisation because  of  overlap between  collinear and soft regions.
Formally such an overlap  leads to the  endpoint singularities in the collinear
integrals.  The well known  example is the nucleon electromagnetic form
factor $F_{2}$ \cite{Belitsky:2002kj}.  However  the short
distance annihilation of $Q\bar{Q}$ pair into  three  gluons in  of $J/\psi $ decay  has different properties.
  None of the three  virtual gluons can have utrasoft momentum  because the coupling of the ultrasoft gluon 
   with the heavy quark  is suppressed  at  least by one additional power of the small velocity $v$.  
   Therefore the annihilation to hard and ultrasoft gluons are suppressed.   This observation  allows one to
conclude  that the helicity suppressed amplitude $A_{2}$ can also be factorised in  the hard and soft  contributions  like the amplitude $A_{1}$.
 Therefore  the amplitude $A_2$ can also be calculated  within the standard collinear  factorisation framework.

\section{Calculation of the amplitude $A_{2}$}
\begin{figure}[ptb]%
\centering
\includegraphics[width=2.0in]{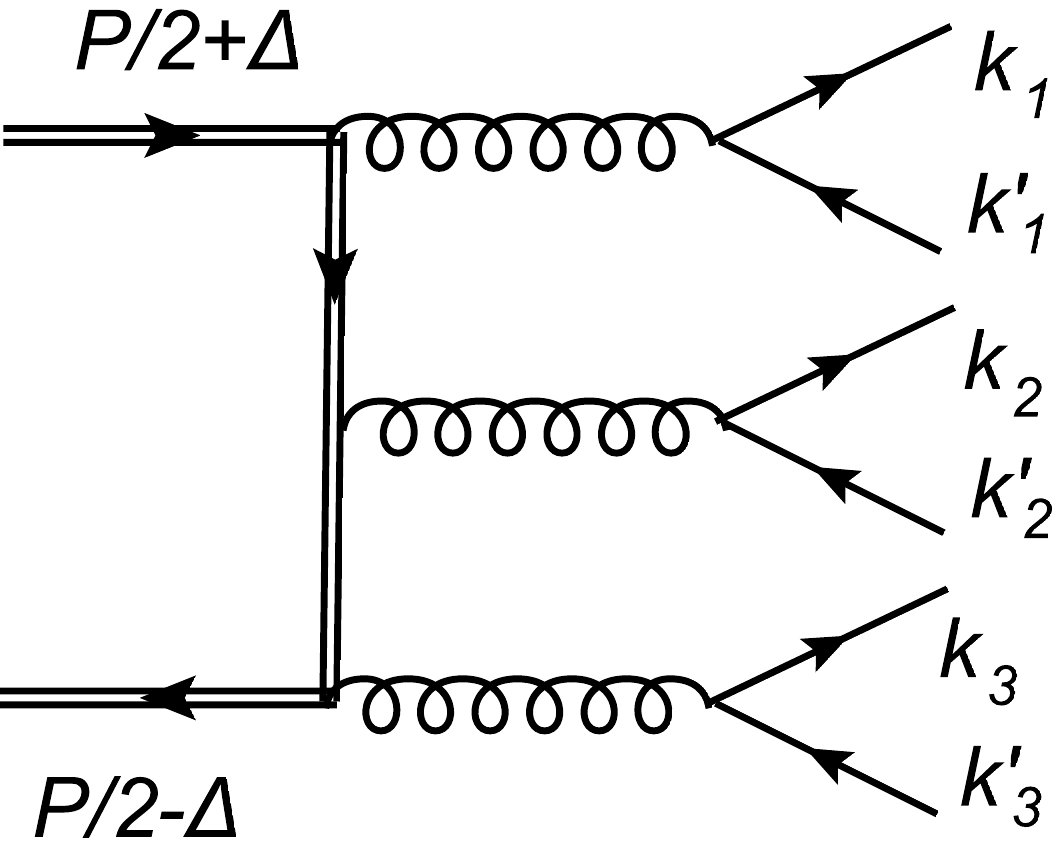}
\caption{One of the six  hard diagrams which contribute to the hard coefficient function.   }
\label{hard_diagram}
\end{figure}

\lab{calc}
The hard coefficient function is given by the hard $Q\bar Q$ annihilation into  three gluons which 
 further creates light quark-antiquark pairs forming the final $p\bar p$ state, the typical diagram is shown in Fig.\ref{hard_diagram}.
The nucleon matrix elements
are described by the operators of twist-3 and twist-4. In present
calculation we take into account only the three-quark operators of twist-4
and neglect the quark-gluon operators. Such approximation is based on the
assumption that the quark-gluon  matrix elements are relatively small, see {\it e.g.} Ref.\cite{Braun:2000kw} .   
The properties of the twist-4 LCDAs were studied in Refs. \cite{Braun:2000kw, Braun:2006hz, Anikin:2013aka, Braun:2008ia, Anikin:2013yoa}.
 For the convenience of the reader we provide  a brief description of these  matrix elements and corresponding LCDAs in
Appendix~\ref{AppA}. 

In general, the calculation of amplitude $A_{2}$ is quite standard. We
compute the diagrams in the momentum space writing the appropriate
projections for the collinear matrix elements. For that purpose we need the
projection for the twist-4   matrix element. As a rule, the projections of
higher twist matrix elements include the derivatives with respect to  quark
transverse momenta. Corresponding formulae can be derived easily from the 
parametrisation of such a matrix element for the corresponding off  light-cone correlator.
The technical problem is that for a nucleon matrix element one has to
consider many various Dirac structures,  that makes the calculation  quite
complicated.  In this work we define the twist-4 projection  related with the three-quark operators which are
built from the large components of the collinear quark fields and transverse
derivatives. Such matrix elements allow one to simplify  the calculations and the final expression for the amplitude. 
Below we briefly discuss the main steps of our calculation. In  order to make simpler  the connection 
with the nucleon matrix elements in Appendix~\ref{AppA},  below we compute the amplitude for the time  reversal 
process $p+\bar{p}\rightarrow J/\psi $ with $(k+k^{\prime})^{2}=M_{\psi }^{2}$.  

In order to perform matching let us  write the expansion for the amplitudes as 
\bea
A_1=A_1^{(0)}+\mathcal{O}(1/m_Q^2), \quad A_2=A_2^{(1)}+\mathcal{O}(1/m_Q^2),  \quad A_2^{(1)}/A_1^{(0)}\sim \mathcal{O}(1/m_Q^2).
\eea
In order  to expand the nucleon spinors in the definition of the  amplitude in Eq.(\ref{def:M}) with respect to powers of $1/M_\psi$. 
we use the equations of the motion 
\bea
(\Dsl{k}-m_N)N(k)=0, \quad \bar{V}(k')(\Dsl{k}'+m_N)=0,
\eea
and perform  decompositions of the spinors on large and small components  
\bea
N_{\bar{n}}=\nbn N(k),\quad   \nnb N(k)=\frac{m_N}{k_{+}}\frac{\Dsl{n}}{2}N_{\bar{n}}, \quad  \bar{V}_n=\bar{V}(k')\nbn, 
\quad  \bar{V}(k')\nnb=-\frac{m_N}{k'_{-}} \bar{V}_n \frac{\Dsl{\bar{n}}}2.
\eea
Using these definitions and performing the expansion of the amplitude $M[p\bar{p}\to J/\psi ]$ one finds  
\bea
 M[p\bar{p}\to J/\psi ]&=&
(\epsilon_{\psi} )_\sigma \bar V_{n}\gamma_{\bot}^\sigma {N}_{\bar{n}}  ~ A^{(0)}_{1} 
\nonumber \\
&+&\frac{1}{4m_{N}}
\left\{  \left(  \epsilon_{\psi}\cdot n\right)
k_{-}^{\prime}~-k_{+}\left(  \epsilon_{\psi}\cdot\bar{n}\right)  \right\}
~\bar V_{n}{N}_{\bar{n}}~\left[  ~A^{(1)}_{2}+\frac{4m_{N}^{2}}{ k_{+}k_{-}^{\prime} }A^{(0)}_{1}\right],
\label{Mexp}
\eea
where we only keep the leading-oder contributions in front of the nucleon bispinors. 
This expression shows that  the coefficient in front of  chiral odd combination $\bar V_{n}{N}_{\bar{n}}$  includes  the spin flip amplitude $A^{(1)}_{2}$
and the kinematical power correction $\sim A^{(0)}_{1}$.  On the other hand, the  expression for this coefficient can be computed in  the EFT  framework. 

The contribution of diagrams in Fig.\ref{hard_diagram} can be  written as 
\begin{eqnarray}
&& i(2\pi )^{4}\delta (p-k-k^{\prime })\,  M[p\bar{p}\to J/\psi ] \simeq \int dz_{1}^{\prime }dz_{2}^{\prime }dz_{3}^{\prime }\int
dz_{1}dz_{2}dz_{3}~D(z_{i}^{\prime },z_{j})
 \notag \\ && 
\quad \quad \times ~\left\langle P\right\vert \bar{Q}_{\alpha }(z_{1}^{\prime })Q_{\beta }(z_{2}^{\prime })\left\vert0\right\rangle 
\left\langle 0\right\vert u_{1}(z_{1})u_{2}(z_{2})d_{3}(z_{3})~\bar{u%
}_{1^{\prime }}(z_{1})\bar{u}_{2^{\prime }}(z_{2})\bar{d}_{3^{\prime
}}(z_{3})\left\vert k,k^{\prime }\right\rangle .  \label{A2v0}
\end{eqnarray}%
Here, the function$~D_{\beta \alpha }(z_{i}^{\prime },z_{j})$ represents the
sum of the hard diagrams in position space. This function depends from 
spinor and colour indices,  which are not shown  for simplicity.  We also imply the factorisation of heavy quark sector and
therefore we write in Eq.(\ref{A2v0})  the  product of the charmonia and
proton-antiproton matrix elements. The light quark operator is constructed
from the $u$- and $d$- quark fields  and we simplify the writing of the
spinor indices 
\begin{equation}
u_{\sigma _{i}}\equiv u_{i},~\ \ \bar{u}_{\sigma _{i}^{\prime }}\equiv \bar{u%
}_{i^{\prime }}\text{. }
\end{equation}
The colour indices are also not shown for simplicity. 

The calculation of the heavy quark matrix element in NRQCD \cite{Bodwin:1994jh}  
is a well known technique, therefore, we skip the details and only write the final  expression 
\begin{equation}
~\left\langle P\right\vert \bar{Q}_{\alpha }(z_{1}^{\prime })Q_{\beta
}(z_{2}^{\prime })\left\vert 0\right\rangle \simeq e^{im_{Q}\omega \cdot
(z_{1}^{\prime }+z_{2}^{\prime })}\frac{1}{4}\left[ (1-\NEG{\omega})\NEG%
{\epsilon}_{\psi }\right] _{\beta \alpha }~f_{\psi },  \label{meQQbar}
\end{equation}%
where the constant $f_{\psi }$ is  defined in Eq.(\ref{def:fpsi}).  

Consider the proton-antiproton matrix element 
\begin{equation}
M_{h}=\left\langle 0\right\vert u_{1}(z_{1})u_{2}(z_{2})d_{3}(z_{3})~\bar{u}%
_{1^{\prime }}(z_{1})\bar{u}_{2^{\prime }}(z_{2})\bar{d}_{3^{\prime
}}(z_{3})\left\vert k,k^{\prime }\right\rangle .  \label{Mhv0}
\end{equation}%
We rewrite the given quark operator as the product of the collinear
operators of twist-3 and twist-4.  For that purpose, we expand the quark
fields into large and small collinear components 
\begin{eqnarray}
\psi (x) =\frac{\nbs\ns}{4}\psi (x)+\frac{\ns\nbs}{4}\psi
(x)=\xi _{\bar{n}}(x)+\eta _{\bar{n}}(x),~ 
\bar{\psi}(x) =\ \bar{\psi}(x)\frac{\nbs\ns}{4}+\bar{\psi}(x)%
\frac{\ns\nbs}{4}=\bar{\xi}_{n}(x)+\bar{\eta}_{n}(x).
\end{eqnarray}
The effective field theory  counting rules imply
\begin{eqnarray}
\bar{\xi}_{n}(x_{+}) \sim \xi _{\bar{n}}(x_{-})\sim \lambda
^{2},~~\partial _{\bot }\xi _{\bar{n}}(x_{-})\sim \partial _{\bot }\bar{\xi}%
_{n}(x_{+})\sim \lambda ^{4}, 
~ \eta _{\bar{n}}(x) \sim \bar{\eta}_{n}(x)\sim \lambda ^{4}.\ 
\end{eqnarray}%
At the next step one has to perform the multipole expansion of the fields. To our accuracy we
need 
\begin{eqnarray}
\psi (x) \simeq \xi _{\bar{n}}(x_{-})+\left( x_{\bot }\partial _{\bot
}\right) \xi _{\bar{n}}(x_{-})+\eta _{\bar{n}}(x_{-}),~
\ \bar{\psi}(x) \simeq \bar{\xi}_{n}(x_{+})+\left( x_{\bot }\partial
_{\bot }\right) \bar{\xi}_{n}(x_{+})+\bar{\eta}_{n}(x_{+}),
 \label{LCexp}
\end{eqnarray}%
where we introduced short notations for the light-cone arguments of the fields%
\begin{equation}
~x_{-}\equiv (x\bar{n})\frac{n}{2},~ x_{+}\equiv (xn)\frac{\bar{n}}{2}.
\end{equation}%
We assume that the collinear and hard fields can be completely decoupled
(factorised) in the effective Lagrangian, which can be written as the sum of
three  contributions corresponding to the hard, $n$- and $\bar{n}$%
-collinear sectors. Such factorisation implies that the leading-order 
collinear gluon fields 
\begin{equation}
\bar{n}\cdot A^{(n)}(x_{+})\sim n\cdot A^{(\bar{n})}(x_{-})\sim \mathcal{O}%
(\lambda ^{0}),
\end{equation}%
are factorised into the collinear Wilson lines,  which ensure the colour gauge
invariance. The detailed discussion of  this  factorisation is quite
complicated and requires a separate consideration. Therefore,  we accept this
fact as a plausible assumption. The factorisation of the hard and collinear
sectors implies that the operator in (\ref{Mhv0}) will be modified by the
redefinition of the collinear fields 
\begin{equation}
\xi _{\bar{n}}(x_{-})\rightarrow W_{\bar{n}}^{\dag }(x_{-})\xi _{\bar{n}%
}(x_{-}),~\ \bar{\xi}_{n}(x_{+})\rightarrow \bar{\xi}_{n}(x_{+})W_{n}(x_{+}),
\end{equation}%
\begin{equation}
\eta _{\bar{n}}(x_{-})\rightarrow W_{\bar{n}}^{\dag }(x_{-})\eta _{\bar{n}%
}(x_{-}),\ \bar{\eta}_{n}(x_{+})\rightarrow \bar{\eta}%
_{n}(x_{+})W_{n}(x_{+}),
\end{equation}%
where the collinear Wilson lines $W_{n,\bar{n}}$ are defined as 
\begin{equation}
W_{n}(x_{+})=\text{P}\exp ig\int_{-\infty }^{0}ds~~\bar{n}\cdot A^{(n)}(x_{+}\bar{n}%
/2+s\bar{n}),
\end{equation}%
\begin{equation}
W_{\bar{n}}(x_{-})=\text{P}\exp ig\int_{-\infty }^{0}ds~~n\cdot A^{(\bar{n}%
)}(x_{-}n/2+sn).
\end{equation}
The terms with the transverse derivatives of the quark fields must be
redefined as 
\begin{eqnarray}
\partial _{\bot }\xi _{\bar{n}}(x_{-}) \rightarrow W_{\bar{n}}^{\dag
}(x_{-})\partial _{\bot }\xi _{\bar{n}}(x_{-}) 
=\left[ W_{\bar{n}}^{\dag }(x_{-})\partial _{\bot }W_{\bar{n}}(x_{-})%
\right] ~W_{\bar{n}}^{\dag }\xi _{\bar{n}}(x_{-})+~\partial _{\bot }W_{\bar{n%
}}^{\dag }\xi _{\bar{n}}(x_{-}),
\end{eqnarray}%
where we assume that the derivative must be applied only inside the square
brackets.  The contribution with$~\left[ W_{\bar{n}%
}^{\dag }\partial _{\bot }W_{\bar{n}}\right] $ is not gauge invariant, but it
can be associated with the quark-gluon operators. This term  must be
combined with the diagrams with emissions of collinear transverse gluons from
the quark lines. The sum of  such contributions gives the  gauge invariant
quark-gluon operators of twist-4. However, we neglect such operators and
therefore we can skip this contribution. Hence, we can write 
\begin{equation}
\partial _{\bot }\xi _{\bar{n}}(x_{-})\rightarrow ~\partial _{\bot }W_{\bar{n%
}}^{\dag }(x_{-})\xi _{\bar{n}}(x_{-}).
\end{equation}%
Notice that this term is already gauge invariant. 

After the factorisation of hard and collinear sectors, the relevant
contributions to the hadronic matrix element (\ref{Mhv0}) can be written as 
\begin{eqnarray}
M_{h} \simeq \left\langle 0\right\vert \left[ O(z_{i+})\right] _{\text{tw3%
}}~\left\vert k^{\prime }\right\rangle ~\left\langle 0\right\vert \left[
O(z_{i-})\right] _{\text{tw4}}\left\vert k\right\rangle +~\left\langle 0\right\vert \left[ O(z_{i+})\right] _{\text{tw4}%
}~\left\vert k^{\prime }\right\rangle ~\left\langle 0\right\vert \left[
O(z_{i-})\right] _{\text{tw3}}\left\vert k\right\rangle ,
\end{eqnarray}%
where the  leading twist-3 operators read
\begin{eqnarray}
\left[ O(z_{i+})\right] _{\text{tw3}} =\bar{\chi}_{1^{\prime }}(z_{1+})%
\bar{\chi}_{2^{\prime }}(z_{2+})\bar{\chi}_{3^{\prime }}(z_{3+}), ~~ 
\left[ O(z_{i-})\right] _{\text{tw3}} =\chi _{1}(z_{1-})\chi
_{2}(z_{2-})\chi _{3}(z_{3-}).
\end{eqnarray}%
In these formulas we used the standard notation for the gauge invariant
blocks%
\begin{equation}
\bar{\chi}(z_{+})=\bar{\xi}_{n}(z_{+})W_{n}(z_{+}),~ \chi (z_{-})=W_{\bar{%
n}}^{\dag }(x_{-})\xi _{\bar{n}}(x_{-}).  \label{chi}
\end{equation}%
We do not show explicitly  the quark flavour assuming that Dirac
indices $1,2$ correspond to the $u$-quarks.  We also do not indicate
 the collinearity ($n$ and $\bar{n}$) of the fields explicitly,  assuming that
the field arguments allows one  to conclude about  the collinear sector.

The obtained  twist-4 operators  include the fields with the
transverse derivatives $\partial _{\bot }\chi $.  For the nucleon case one
finds 
\begin{align}
&\left[ O(z_{i-})\right] _{\text{tw4}} =-i\left( z_{1}-z_{3}\right) _{\bot
}^{\alpha }\left[ i\partial _{\bot\alpha }\chi _{1}(z_{1-})\right] \chi
_{2}(z_{2-})\chi _{3}(z_{3-}) 
\notag \\ 
&-i\left( z_{2}-z_{3}\right) _{\bot }^{\alpha }\chi _{1}(z_{1-})\left[
~i\partial _{\bot \alpha }\chi _{2}(z_{2-})\right] \chi _{3}(z_{3-})
-\frac{1}{2}\left[ (in\partial )^{-1}\NEG{n}i\NEG{\partial}_{\bot }\chi
(z_{1-})\right] _{1}~\chi _{2}(z_{2-})~\chi _{3}(z_{3-}) 
\notag \\ 
&-\frac{1}{2}\chi _{1}(z_{1-})~\left[ (in\partial )^{-1}\NEG{n}i\NEG
{\partial}_{\bot }\chi (z_{2-})\right] _{2}~\chi_{3}(z_{3-})
+\frac{1}{2}[{i\partial}_{\bot\alpha } \chi _{1}(z_{1-})]~\chi _{2}(z_{2-})~
\left[ (in\partial )^{-1}\NEG{n}\gamma^\alpha \chi (z_{3-})\right] _{3}
\nonumber \\
&+\frac{1}{2}\chi _{1}(z_{1-})~[{i\partial}_{\bot \alpha}\chi _{2}(z_{2-})] \left[ (in\partial )^{-1}\NEG%
{n}\gamma^\alpha\chi (z_{3-})\right] _{3}. 
 \label{Otw4}
\end{align}%
 The first two terms in {\it rhs} of Eq.(\ref{Otw4}) appear from to the multipole  expansion of the fields (this was explained above). 
 The remnant terms in {\it rhs} of Eq.(\ref{Otw4})  appear from the small collinear components $\eta (z_{i})$. 
 The latter can be  rewritten  using QCD EOM 
\begin{align}
&W_{\bar{n}}^{\dag }(x_{-})\eta _{\bar{n}}(x_{-})=-W_{\bar{n}}^{\dag }(x_{-})%
\frac{\NEG{n}}{2}(inD)^{-1}i\NEG{D}_{\bot }\xi _{\bar{n}}(x_{-})
=-\frac{\NEG{n}}{2}(in\partial )^{-1}W_{\bar{n}}^{\dag }(x_{-})i\NEG%
{D}_{\bot }\xi _{\bar{n}}(x_{-})
\notag \\ &
=-\frac{\NEG{n}}{2}(in\partial )^{-1}\left[ W_{\bar{n}}^{\dag }(x_{-})i\NEG%
{D}_{\bot }W_{\bar{n}}(x_{-})\right] \chi _{\bar{n}}(x_{-})
-\frac{\NEG{n}}{2}%
(in\partial )^{-1}i\NEG{\partial}_{\bot }\chi _{\bar{n}}(x_{-})
\notag \\ 
& \simeq -\frac{\NEG{n}}{2}(in\partial )^{-1}i\NEG{\partial}_{\bot }\chi _{\bar{%
n}}(x_{-}), 
 \label{etaEOM}
\end{align}%
where we again neglected the contribution with $\left[ W_{\bar{n}}^{\dag
}(x_{-})i\NEG{D}_{\bot }W_{\bar{n}}(x_{-})\right]$, which gives the
quark-gluon operator. We also used that the matrix element of any  operator with the total transverse derivative  
vanishes $\langle0\vert \partial_\bot \mathcal{O}\vert k\rangle=0$ and therefore can be neglected. This gives
\begin{align}
\chi _{1}(z_{1-})~\chi _{2}(z_{2-})~\partial _{\bot }^{\alpha }\chi
_{3}(z_{3-}) \simeq -\left[ \partial _{\bot }^{\alpha }\chi _{1}(z_{1-})%
\right] \chi _{2}(z_{2-})~\chi _{3}(z_{3-}) 
\notag \\
-\chi _{1}(z_{1-})~\left[ \partial _{\bot }^{\alpha }\chi _{2}(z_{2-})%
\right] ~\chi _{3}(z_{3-}).
\end{align}%

The matrix element of twist-3 light-cone operator is given in Eq.(\ref{tw3me}).
The expression for the twist-4 matrix element is more complicated. Using the
Fierz identities%
\begin{align}
\chi _{1}(z_{1-})\chi _{2}(z_{2-})& =-\frac{1}{8}\left[ \nbs C\right]
_{12}~\chi (z_{1-})C\Dsl{n}\chi (z_{2-})
 -\frac{1}{8}\left[\nbs \gamma _{5}C\right] _{12}~\chi (z_{1-})C\NEG%
{n}\gamma _{5}\chi (z_{2-})  \notag \\
&+\frac{1}{8}\left[\nbs \gamma _{\bot }^{\sigma }C\right] _{12}~\chi
(z_{1-})C\NEG{n}\gamma _{\bot \sigma}\chi (z_{2-}). 
 \label{FVAT}
\end{align}%
we rewrite this operator as a sum 
\begin{equation}
\left[ O(z_{i-})\right] _{\text{tw4}%
}=O_{V}(z_{i-})+O_{A}(z_{i-})+O_{T}(z_{i-}),  \label{OVAT}
\end{equation}%
where the operators $O_{V}$, $ O_{A}$ and $O_{T}$  correspond to the three
projections in Eq.(\ref{FVAT}), respectively. Consider the operator $O_{V}$.
It can be written as a sum of two terms,  which only include $\partial _{\bot }^{\alpha
}\chi _{1}(z_{1-})$  or  $\partial _{\bot }^{\alpha }\chi _{2}(z_{2-})$,
respectively 
\begin{equation}
O_{V}(z_{i-})=O_{V1}(z_{i-})+O_{V2}(z_{i-}).
\end{equation}%
From Eq.(\ref{Otw4}) one finds 
\begin{align}
O_{V1}(z_{i-})=\frac{1}{8}\left[ \nbs C\right] _{12}i\left(
z_{1}-z_{3}\right) _{\bot }^{\alpha }\left[ i\partial _{\bot \alpha}\chi
(z_{1-})\right] C\ns\chi (z_{2-})~\chi _{3}(z_{3-})
\notag\\
+\frac{1}{8}\left[ \ns\gamma _{\bot }^{\alpha }\nbs C\right] _{12}\frac{1}{2}%
\left[ (in\partial )^{-1}i\partial _{\bot \alpha }\chi (z_{1-})\right] C\ns \chi (z_{2-})~\chi _{3}(z_{3-})
\notag \\
-\frac{1}{8}\left[ \nbs C\right] _{12}\frac{1}{2}\left[ i\partial _{\bot\alpha
}\chi (z_{1-})\right] C\NEG{n}\chi (z_{2-})~\left[ (i\bar{n}%
\partial )^{-1}\ns\gamma _{\bot }^{\alpha }\chi (z_{3}^{+})\right] _{3}.
\label{OV1}
\end{align}%
The matrix element of this operator can be easily computed with the help of  Eq.(\ref{meV1cal}). This gives
\begin{equation}
~\left\langle 0\right\vert \left[ (in\partial )^{-1}i\partial _{\bot \alpha
}\chi (z_{1-})\right] C\Dsl{n}\chi (z_{2-})\chi _{3}(z_{3-})\left\vert
k\right\rangle
=m_{N}~\left[ \gamma _{\bot\alpha }\gamma _{5}N_{\bar n}\right] _{3}\text{FT}%
\left[ \frac{1}{x_{1}}\mathcal{V}_{1}(x_{i})\right] ,
\end{equation}%
\begin{align}
&\left\langle 0\right\vert \left[ i\partial _{\bot }^{\alpha }\chi (z_{1-})\right] C\Dsl{n}\chi (z_{2-})~\left[ (i\bar{n}\partial )^{-1}n\gamma _{\bot
\alpha }\chi (z_{3-})\right] _{3}\left\vert k\right\rangle
\notag \\
&\mskip 200mu =m_{N}~\left[ \Dsl{n}\gamma _{\bot }^{\alpha }\gamma _{\bot\alpha }\gamma _{5}N_{\bar n}
\right] _{3}~\left[ \Dsl{\bar{n}}C\right] _{12}\text{FT}\left[ \frac{1}{x_{3}}
\mathcal{V}_{1}(x_{i})\right] ,
\end{align}
where symbol  FT  denotes the Fourier transformation. 
With the help of  these equations  we obtain
\begin{align}
\left\langle 0\right\vert & O_{V1}(z_{i-})\left\vert k\right\rangle =k_{+}m_N
\frac{1}{8}\left[ \nbs C\right] _{12}\left[ \gamma _{\bot }^{\alpha
}\gamma _{5}N_{\bar n}\right] _{3}~i\left( z_{1}-z_{3}\right) _{\bot \alpha}~
\text{FT}\left[ \mathcal{V}_{1}(x_{i})\right]
\notag \\
&\mskip 100mu +\frac{1}{16}m_N\left[ \gamma _{\bot }^{\alpha }\gamma _{5}N_{\bar n} \right]
_{3}~\left[ \ns \gamma _{\bot \alpha}\nbs C\right] _{12}\text{FT}\left[ 
\frac{1}{x_{1}}\mathcal{V}_{1}(x_{i})\right]
\notag \\ & \mskip 150mu
-\frac{1}{16}m_N~\left[ \ns \gamma _{\bot }^{\alpha }\gamma
_{\bot\alpha }\gamma _{5}N_{\bar n} \right] _{3}~\left[ \nbs C\right] _{12}\text{
FT}\left[ \frac{1}{x_{3}}\mathcal{V}_{1}(x_{i})\right] .  \label{meOV1}
\end{align}%
Now we have the  all required matrix elements (\ref{meQQbar}),
(\ref{tw3me}) and (\ref{meOV1}), which we need in order to calculate the  contribution to the amplitude $M$
\begin{align}
&& i(2\pi )^{4}\delta (p-k-k^{\prime })\, M[\left[ O(z_{i+})\right] _{\text{tw3}}, O_{V1}(z_{i-})] 
=\int dz_{1}^{\prime }dz_{2}^{\prime }dz_{3}^{\prime }\int
dz_{1}dz_{2}dz_{3}~D(z_{i}^{\prime },z_{j})~\notag \\
&&\times \left\langle P\right\vert \bar{Q}%
_{\alpha }(z_{1}^{\prime })Q_{\beta }(z_{2}^{\prime })\left\vert
0\right\rangle \left\langle 0\left\vert 
\left[ O(z_{i+})\right] _{\text{tw3}}\right\vert k^{\prime }\right\rangle
\left\langle 0\right\vert O_{V1}(z_{i-})\left\vert k\right\rangle .
\end{align}%
One can use the fact  that all our diagrams have the same structure with respect to
spinor indices 
\begin{equation}
\left[ D(z_{i}^{\prime },z_{j})\right] _{123;1^{\prime }2^{\prime }3^{\prime
}}=D^{\mu _{1}\mu _{2}\mu _{3}}(z_{i}^{\prime },z_{j})~\left[ \gamma _{\mu
_{1}}\right] _{1^{\prime }1}\left[ \gamma _{\mu _{2}}\right] _{2^{\prime }2}%
\left[ \gamma _{\mu _{3}}\right] _{3^{\prime }3},
\end{equation}%
where the $\gamma $-matrices $\gamma _{\mu _{i}}$ originate from the light quark-gluon
vertices in the diagrams as in Fig.\ref{hard_diagram}. Substitution of the matrix
elements from (\ref{meQQbar}), (\ref{tw3me}) and (\ref{meOV1}) and
contractions of the spinor indices yields
\begin{align}
&& i(2\pi )^{4}\delta (p-k-k^{\prime })\, M[V_{1},\mathcal{V}_{1}]=\int dz_{1}^{\prime }dz_{2}^{\prime
}dz_{3}^{\prime }\int dz_{1}dz_{2}dz_{3}~e^{im_{Q}\omega \cdot
(z_{1}^{\prime }+z_{2}^{\prime })}
\notag \\
&&
 \times  
 \int Dy_{i}~V_{1}(y_{i})~e^{-i(k_{1}^{\prime }z_{1})-i(k_{2}^{\prime }z_{2})-i(k_{3}^{\prime }z_{3})}\,
 T _{3g\to p\bar{p}}[ \mathcal{V}_{1} ]~f_{\psi }\frac{1}{4}\text{Tr}\left[
(1-\NEG{\omega})\NEG{\epsilon}_{\psi }D^{\mu _{1}\mu _{2}\mu
_{3}}(z_{i}^{\prime },z_{j})\right] ,  \label{A2v2}
\end{align}%
with 
\begin{align}
 &T _{3g\to p\bar{p}}[\mathcal{V}_{1}]=-\frac{1}{4}m_{N}~\bar{V}_n\gamma _{5}\gamma
_{\mu_3}\gamma _{\bot }^{\alpha }\gamma _{5}N_{\bar n}~\frac{1}{4}\text{Tr}\left[ \gamma
_{\mu_1}\Dsl{k}C\left( C\Dsl{k}^{\prime }\gamma _{\mu_2}\right) ^{\top }\right] 
\frac{\partial }{\partial k_{1\bot }^{\alpha }}\text{FT}\left[ \mathcal{V}%
_{1}(x_{i})\right]
\notag\\
&+\frac{m_{N}}{16}\bar{V}_n\gamma _{5}\gamma _{\mu _{3}}\gamma _{\bot }^{\alpha
}\gamma _{5}N _{\bar n}\frac{1}{4}\text{Tr}\left[ \gamma _{\mu _{1}}\ns\gamma _{\bot\alpha
}\nbs C\left( C\NEG{k}^{\prime }\gamma _{\mu _{2}}\right) ^{\top
}\right] \text{FT}\left[ \frac{1}{x_{1}}\mathcal{V}_{1}(x_{i})\right]
\notag\\
&-\frac{m_{N}}{16}\bar{V}_n\gamma _{5}\gamma _{\mu _{3}}\Dsl{n}\gamma _{\bot
}^{\alpha }\gamma _{\bot \alpha}\gamma _{5}N_{\bar n}\frac{1}{4}\text{Tr}\left[
\gamma _{\mu _{1}}\nbs C\left( C\Dsl{k}^{\prime }\gamma _{\mu _{2}}\right)
^{\top }\right] \text{FT}\left[ \frac{1}{x_{3}}\mathcal{V}_{1}(x_{i})\right] .
 \label{()pp}
\end{align}%
For simplicity we show in this expression only the contribution with the
twist-3 DA $V_{1}(y_{i})$.

The term with the transverse derivative $\partial
/\partial k_{1\bot }^{\alpha }$ in (\ref{()pp}) arises from the contribution
with $\ i\left( z_{1}-z_{3}\right) _{\bot }^{\alpha }$ in  Eq.(\ref{meOV1}%
).  In order to  describe  the transition  $i(z_1-z_2)_\perp\to \partial/\partial k_{1\bot }$,
we used that the corresponding contribution can be written as 
\begin{equation}
i(z_1-z_3)_\perp \text{FT}\left[ \mathcal{V}
_{1}(x_{i})\right] =\left. -\frac{\partial }{\partial k_{1\bot }^{\alpha }}
\int Dx_{i}~e^{-i(k_{1}z_{1})-i(k_{2}z_{2})-i(k_{3}z_{3})}\mathcal{V}
_{1}(x_{i})\right\vert _{k_{\bot i}=0},
\end{equation}%
where partonic momenta $k_i$ have the transverse components
\begin{equation}
k_{1}=x_{1}k+k_{1\bot },~~k_{2}=x_{2}k+k_{2\bot },~\ k_{3}=x_{3}k-k_{1\bot
}-k_{2\bot }, ~~ k\simeq k_+ \frac{\bar n}{2} .\label{k:def}
\end{equation}

Now we can perform the integrations over $dz_{i}^{\prime }$ and $dz_{j}$ in 
(\ref{A2v2}). This corresponds to the Fourier transformation of
the diagrams $D^{\mu _{1}\mu _{2}\mu _{3}}(z_{i}^{\prime },z_{j})$   to
the momentum space
\begin{equation}
D^{\mu _{1}\mu _{2}\mu _{3}}(z_{i}^{\prime },z_{j})~\rightarrow D^{\mu
_{1}\mu _{2}\mu _{3}}(k_{i}^{\prime },k_{j}).
\end{equation}
  This gives
\begin{equation}
i M[V_{1}\mathcal{V}_{1}]=f_{\psi }\int Dy_{i}~V_{1}(y_{i})\int Dx_{i}~%
\mathcal{V}_{1}(x_{i})~
 \left. \hat T _{3g\to p\bar{p}} ~\frac{1}{4}\text{Tr}\left[ (1-\NEG%
{\omega})\NEG{\epsilon}_{\psi }D^{\mu _{1}\mu _{2}\mu _{3}}(k_{i}^{\prime
},k_{j})\right] \right\vert _{k_{\bot i}=0},
\end{equation}%
where
\begin{eqnarray}
 \hat T _{3g\to p\bar{p}} &=&-\frac{1}{4}m_{N}~\bar{V}_n\gamma _{5}\gamma
_{\mu_3}\gamma _{\bot }^{\alpha }\gamma _{5}N_{\bar n}~\frac{1}{4}\text{Tr}\left[ \gamma
_{\mu_1}\NEG{k}C\left( C\NEG{k}^{\prime }\gamma _{\mu_2}\right) ^{\top }\right] 
\frac{\partial }{\partial k_{1\bot }^{\alpha }}
\notag \\
&&+\frac{1}{x_{1}}\frac{m_{N}}{16}\bar{V}_n\gamma _{5}\gamma _{\mu _{3}}\gamma
_{\bot }^{\alpha }\gamma _{5}N_{\bar n}\frac{1}{4}\text{Tr}\left[ \gamma _{\mu
_{1}}\ns \gamma _{\bot \alpha}\nbs C\left( C\NEG{k}^{\prime }\gamma _{\mu
_{2}}\right) ^{\top }\right]
\notag \\
&&-\frac{1}{x_{3}}\frac{m_{N}}{16}\bar{V}_n\gamma _{5}\gamma _{\mu _{3}}\NEG%
{n}\gamma _{\bot }^{\alpha }\gamma _{\bot\alpha }\gamma _{5}N_{\bar n}\frac{1}{4}%
\text{Tr}\left[ \gamma _{\mu _{1}}\nbs C\left( C\NEG{k}^{\prime }\gamma
_{\mu _{2}}\right) ^{\top }\right] .
\end{eqnarray}%
The analytical expression in $D^{\mu _{1}\mu _{2}\mu _{3}}(k_{i}^{\prime },k_{j})$
describes the contributions of the heavy quark line with the gluon vertices
(with the indices $\mu _{i}$) and the three gluon propagators as in the diagrams in Fig.
\ref{hard_diagram}.  The light quark momenta defined as $k_{i}^{\prime
}=y_{i}k^{\prime }_- n/2$ and as in Eq.(\ref{k:def}).  The similar contributions  must be also
 obtained  for the  other operators in Eq.(\ref{OVAT}).  

Performing the calculations of the traces and comparing with Eq.(\ref{Mexp}) one obtains the final expression
for the required amplitude 
\begin{equation}
A^{(0)}_{2}=\frac{m_{N}^{2}}{M_{\psi }^{2}}\frac{f_{\psi }}{~m_{Q}^{2}}\frac{%
~f_{N}^{~2}}{m_{Q}^{4}}~(\pi \alpha _{s})^{3}\frac{10}{81}~J-\frac{4m_{N}^{2}}{ M_\psi^2}A^{(0)}_{1} ,  \label{A2:res}
\end{equation}%
where the dimensionless convolution integral $J$ reads%
\begin{eqnarray}
J =\frac 14\left( J_{1}[V_{1},\mathcal{V}%
_{i}]+J_{2}[A_{1},\mathcal{V}_{i}]+J_{3}[V_{1},\mathcal{A}_{i}]+J_{4}[A_{1},%
\mathcal{A}_{i}]+ J_{5}[T_{1} ,\mathcal{T}_{ij}]\right) .  
\label{J12345}
\end{eqnarray}%
In square brackets we show the LCDAs, which enter in the integrands. Each
integral in Eq.(\ref{J12345}) is given by 
\begin{align}
J_{n}[X_{1},Y_i]=\int Dy_{i}~\frac{X_{1}(y_{i})}{y_{1}y_{2}y_{3}}%
\int Dx_{i}\mathcal{~}\frac{1}{x_{1}x_{2}x_{3}}
\left\{ \frac{K_{n}(Y_i; x_{i},y_{i})}{D_{1}D_{3}}+\frac{L_{n}(Y_i; x_{i},y_{i})%
}{D_{1}D_{2}}+\frac{N_{n}(Y_i; x_{i},y_{i})}{D_{2}D_{3}}\right\}, 
\label{JiKLN}
\end{align}%
where
\begin{equation}
D_{i}=x_{i}(1-y_{i})+(1-x_{i})y_{i}.
\end{equation}%
The  factors $D_i$ appear from the heavy quark propagators. It is easy to
understand that the three groups in Eq.(\ref{JiKLN}) are related with the three
groups of the diagrams which have appropriate configurations of the heavy quark
momenta. The analytical expressions for the coefficients $\{K_{n},L_{n},N_{n}\}$ read%
\begin{align}
K_{1}(\mathcal{V}_{i}; x_{i},y_{i}) =&\mathcal{V}_{1}(x_{i})\left( \frac{y_{3}-x_{3}}{x_{1}}%
+2\frac{x_{1}-y_{1}}{x_{3}}+2x_{1}+2y_{1}-2y_{3}\right) 
\notag \\
+&\mathcal{V}_{2}(x_{i})~2\left( \frac{x_{1}-y_{1}}{x_{3}}+\frac{x_{1}x_{3}%
}{x_{2}}+x_{1}+y_{1}\right) ,  \label{K1}
\end{align}%
\begin{align}
L_{1}(\mathcal{V}_{i}; x_{i},y_{i}) =\mathcal{V}_{1}(x_{i})\left( \frac{y_{2}-x_{2}}{x_{1}}-%
\frac{4x_{1}x_{2}}{x_{3}}-2x_{2}\right) 
+\mathcal{V}_{2}(x_{i})\left( \frac{y_{1}-x_{1}}{x_{2}}-\frac{4x_{1}x_{2}}{%
x_{3}}-2x_{1}\right) ,
\end{align}%
\begin{align}
&N_{1}(\mathcal{V}_{i};x_{i},y_{i}) =\mathcal{V}_{1}(x_{i})~2\left( \frac{x_{2}-y_{2}}{x_{3}}+\frac{x_{2}x_{3}}{x_{1}}%
+x_2+y_{2}\right) 
\notag \\
&\mskip 100mu +\mathcal{V}_{2}(x_{i})\left( \frac{y_{3}-x_{3}}{x_{2}}+2\frac{x_{2}-y_{2}%
}{x_{3}}+2x_2+2y_{2}-2y_{3}\right) ,
\end{align}%
\begin{align}
K_{2}(\mathcal{V}_{i}; x_{i},y_{i}) =\mathcal{V}_{1}(x_{i})\left( \frac{x_{3}-y_{3}}{x_{1}}
+2x_{1}-2y_{1}+2y_{3}\right) 
+\mathcal{V}_{2}(x_{i})~2\left( x_{1}-y_{1}+\frac{x_{1}x_{3}}{x_{2}}%
\right) ,
\end{align}%
\begin{align}
L_{2}(\mathcal{V}_{i}; x_{i},y_{i}) =\mathcal{V}_{1}(x_{i})\left( \frac{x_{2}-y_{2}}{x_{1}}%
+2x_{2}\right)  +\mathcal{V}_{2}(x_{i})\left( \frac{y_{1}-x_{1}}{%
x_{2}}-2x_{1}\right) ,
\end{align}
\begin{equation}
N_{2}(\mathcal{V}_{i}; x_{i},y_{i})=\mathcal{V}_{1}(x_{i})2 \left( y_2-x_2 -\frac{x_{2}x_{3}}{x_{1}}%
\right) +\mathcal{V}_{2}(x_{i})\left( \frac{y_{3}-x_{3}}{x_{2}}-2x_2+2y_2
-2y_{3}\right) ,
\end{equation}
\begin{align}
K_{3}(\mathcal{A}_{i}; x_{i},y_{i}) =\mathcal{A}_{1}(x_{i})\left( \frac{y_{3}-x_{3}}{x_{1}}-2x_{1}+2y_{1}-2y_{3}\right) 
+\mathcal{A}_{2}(x_{i})~2\left( y_{1}-x_{1}-\frac{x_{1}x_{3}}{x_{2}}\right),
\end{align}%
\begin{align}
L_{3}(\mathcal{A}_{i}; x_{i},y_{i}) =\mathcal{A}_{1}(x_{i})\left( \frac{y_{2}-x_{2}}{x_{1}}-2x_2\right) 
+\mathcal{A}_{2}(x_{i})\left( \frac{x_{1}-y_{1}}{x_{2}}+2x_1\right) ,
\end{align}
\begin{equation}
N_{3}(\mathcal{A}_{i}; x_{i},y_{i})=\mathcal{A}_{1}(x_{i}) 2\left( x_2-y_2+\frac{x_{2}x_{3}}{x_{1}}\right) 
+\mathcal{A}_{2}(x_{i})\left( \frac{x_{3}-y_{3}}{x_{2}}+2x_2-2y_2
+2y_{3}\right) ,
\end{equation}
\begin{eqnarray}
K_{4}(\mathcal{A}_{i}; x_{i},y_{i}) &=&\mathcal{A}_{1}(x_{i})
\left( \frac{x_{3}-y_{3}}{x_{1}}+2\frac{y_{1}-x_{1}}{x_{3}}+2y_{3}-2y_{1}-2x_{1}\right)  
\notag \\
&&+\mathcal{A}_{2}(x_{i})
2\left( \frac{y_{1}-x_{1}}{x_{3}}-\frac{x_{1}x_{3}}{x_{2}}-x_{1}-y_{1}\right) ,
\end{eqnarray}%
\begin{align}
L_{4}(\mathcal{A}_{i}; x_{i},y_{i}) =\mathcal{A}_{1}(x_{i})
\left( \frac{x_{2}-y_{2}}{x_{1}}+4\frac{x_{1}x_{2}}{x_{3}}+2x_{2}\right) 
+\mathcal{A}_{2}(x_{i})
\left( \frac{x_{1}-y_{1}}{x_{2}}+4\frac{x_{1}x_{2}}{x_{3}}+2x_{1}\right) ,
\end{align}%
\begin{eqnarray}
N_{4}(\mathcal{A}_{i}; x_{i},y_{i}) &=&\mathcal{A}_{1}(x_{i})
~2\left( \frac{y_{2}-x_{2}}{x_{3}}-\frac{x_{2}x_{3}}{x_{1}}-y_{2}-x_2\right)
\notag  \\
&&+\mathcal{A}_{2}(x_{i})
\left( \frac{x_{3}-y_{3}}{x_{2}}+2\frac{y_{2}-x_{2}}{x_{3}}+2y_{3}-2y_{2}-2x_2\right) ,
\end{eqnarray}
\begin{align}
K_{5}(\mathcal{T}_{ij}; x_{i},y_{i}) =\left( \mathcal{T}_{21}-\mathcal{T}_{41}\right)
(x_{i})~2\left( \frac{x_{3}-y_{3}}{x_{1}}+2x_{3}\right) 
+\left( \mathcal{T}_{22}-\mathcal{T}_{42}\right) (x_{i})\left( -\frac{%
4x_{1}x_{3}}{x_{2}}\right) ,
\end{align}
\begin{align}
L_{5}(\mathcal{T}_{ij};  x_{i},y_{i}) =\left( \mathcal{T}_{21}-\mathcal{T}_{41}\right)
(x_{i})~2\left( \frac{x_{2}-y_{2}}{x_{1}}+2y_{2}\right) 
+\left( \mathcal{T}_{22}-\mathcal{T}_{42}\right) (x_{i})~2\left( \frac{%
x_{1}-y_{1}}{x_{2}}+2y_{1}\right) ,
\end{align}
\begin{align}
N_{5}(\mathcal{T}_{ij}; x_{i},y_{i}) =\left( \mathcal{T}_{21}-\mathcal{T}_{41}\right)
(x_{i})~\left( -\frac{4x_{2}x_{3}}{x_{1}}  \right) 
+\left( \mathcal{T}_{22}-\mathcal{T}_{42}\right) (x_{i})
2\left(  \frac{x_{3}-y_{3}}{x_{2}}+2x_{3}  \right) .  \label{N5}
\end{align}
The various LCDAs which enter in the expressions in Eqs.(\ref{K1})-(\ref{N5}) are discussed in Appendix \ref{AppA}.  
Notice that the contributions with DAs $\mathcal{V}_2$, $\mathcal{A}_2$ and $\mathcal{T}_{22}-\mathcal{T}_{42}$  
can be reduced to the contributions with $\mathcal{V}_1$, $\mathcal{A}_1$ and $\mathcal{T}_{21}-\mathcal{T}_{41}$
 with the help of the symmetry relations,  see  Eqs.(\ref{VA12tw4}), (\ref{VA12}) and (\ref{T42mnT22}).
The formulas (\ref{A2:res})-(\ref{N5}) represent the  main  result of this work. 

Let us  shortly discuss the properties of the obtained convolution integrals.
These integrals are well defined, i.e they  do not have singularities from the
endpoint regions as it was expected. In order to see this, consider, for instance, the integral
with $V_{1}(y_{i})\mathcal{V}_{i}(x_{i})$ LCDAs. The properties of these functions allow one  to rewrite them as  
\begin{equation}
V_{1}(y_{i})=120y_{1}y_{2}y_{3}\bar{V}(y_{i}), ~~
\mathcal{V}_{i}(x_{i})=3x_{1}x_{2}x_{3}\mathcal{\bar{V}}_{i}(x_{i}),
\label{VC1bar}
\end{equation}%
where the functions $\bar{V}(y_{i})$ and $\mathcal{\bar{V}}_{i}(x_{i})$ are
some nonsingular functions when  momentum fractions are small $y_i, x_i\sim 0$. For the
twist-4 DA $\mathcal{V}_{i}(x_{i})$ this follows from Eq.(\ref{Vical}). 
Substituting (\ref{VC1bar}) into the convolution integral,
we obtain 
\begin{equation}
J_{1}[V_{1},\mathcal{V}_{i}]=360\int Dy_{i}~\bar{V}(y_{i})\int Dx_{i}\left( 
\frac{\bar{K}_{i}(x_{i},y_{i})}{D_{1}D_{3}}+\frac{\bar{L}_{i}(x_{i},y_{i})}{%
D_{1}D_{2}}+\frac{\bar{N}_{i}(x_{i},y_{i})}{D_{2}D_{3}}\right),
\label{J1bar}
\end{equation}%
where the coefficients $\bar{K}_{i},\bar{L}_{i}$ and $\bar{N}_{i}$ depend on 
$\mathcal{\bar{V}}_{i}(x_{i})$. The endpoint singularities can be produced
by the most singular contributions in these coefficients with the factors $
1/x_{i}$. Consider for instance the terms with $1/x_{1}$ which are dangerous
in the limit $x_{1}\rightarrow 0$. The most singular terms with $1/x_1$ give
\begin{align}
\frac{\bar{K}_{i}(x_{i},y_{i})}{D_{1}D_{3}}+\frac{\bar{L}_{i}(x_{i},y_{i})
}{D_{1}D_{2}}+\frac{\bar{N}_{i}(x_{i},y_{i})}{D_{2}D_{3}} 
=\frac{\mathcal{\bar{V}}_{1}(x_{i})}{x_{1}}\left( \frac{y_{3}-x_{3}}{
D_{1}D_{3}}+\frac{y_{2}-x_{2}}{D_{1}D_{2}}+\frac{2x_{2}x_{3}}{D_{2}D_{3}}
\right) +~\ldots 
\end{align}%
\begin{equation}
=2\mathcal{\bar{V}}_{1}(x_{i})\frac{x_{2}x_{3}(1-2y_{1})+y_{2}y_{3}}{
D_{1}D_{2}D_{3}}+~\ldots =F(x_1, x_2,y_i),
\end{equation}%
where for simplicity dots  denote the  contributions without $1/x_1$. 
Hence, we see that the  dangerous singularity $1/x_{1}$ cancel. The resulting expression
is power suppressed in the endpoint region $x_{1}\sim 0$ ($\eta \ll1$) 

\begin{align}
J_{1}[V_{1},\mathcal{V}_{i}]_{us}\sim \int Dy_{i}~\bar{V}(y_{i})~\int_{0}^{\eta}dx_{1} \int_{0}^{1}dx_{2}\mathcal{\bar{V}}_{1}(0,x_{2},\bar{x}_{2}) F(0, x_{2},y_{i}) 
\notag \\
\sim \eta \int Dy_{i}~\bar{V}(y_{i})~\int_{0}^{1}dx_{2}\mathcal{\bar{V}}%
_{1}(0,x_{2},\bar{x}_{2})F(0, x_{2},y_{i}),
\end{align}
The cut-off parameter $\eta $
is small and can be associated with the power suppressed  scale $\eta \sim
\lambda ^{2}$.  We see that the integral over  small fraction $x_{1}$ in
the endpoint limit $x_{1}\sim \eta $  is power suppressed.  The function 
$F(0, x_{2},y_{i})$ does not have any power singularities in the other endpoint regions 
 and therefore  the  total integral is of order $\eta$. This
means that the contributions from the endpoint regions are power suppressed
and the integrals over the momentum fractions are well defined. The same
conclusions are also true for the other  contributions with $1/x_{2}$ and $1/x_{3} $
and for the integrals with other combinations of LCDAs. 

The absence of the endpoint divergencies   is closely
related with the suppression of the ultrasoft gluons with momentum $k_\mu\sim m_Q v^2$ in the heavy quark annihilation.
One can  show without the explicit calculation that the singular terms, arising due to  the ultrasoft
gluon in the individual diagrams, will  cancel in the sum of all diagrams. This consideration is briefly discussed
in Appendix \ref{AppB}.  The  cancellation of the endpoint singularities provides a good
check of the  obtained  expressions for the hard coefficient functions.

\section{Phenomenology}
\lab{phen}
In this section the obtained amplitude $A_{2}$ is used for a qualitative
analysis of the branching ratio  and  angular behaviour of the cross
section $e^{+}e^{-}\rightarrow J/\psi \rightarrow p\bar{p}$. Except of the 
corrections, associated with the higher  Fock components of  hadronic wave
functions, there are also  relativistic corrections associated with the 
charmonium wave function.  These corrections are formally suppressed by
the power of $v^{2}$,  at the same time the hard annihilation mechanism is
strongly suppressed by  power(s) of the small $\alpha _{s}$. In  
the Coulomb limit    $m_{Q}v^{2}\gg \Lambda $   the colour-octet
contribution can be described by the  annihilation with  one  hard  and
two ultrasoft  gluons. Such contribution is associated with the colour-octet operator  and  referred  as the octet  contribution. 
  This contribution is obviously of order $\alpha _{s}(m_{c}^{2})$.
We assume  that $\alpha _{s}((m_{Q}v^{2})^{2}) \gg \alpha _{s}(m_{c}^{2})$ and that it can be estimated  to be  of order one for the real world.  
Then a  simple estimate for the decay amplitudes gives  
\begin{equation}
\frac{A_{oct}}{A_{sing}}\sim v^{2}\alpha _{s}/\alpha _{s}^{3}\sim
v^{2}/\alpha _{s}^{2},
\end{equation}%
where  all $\alpha _{s}$ are defined at  the hard scale $\sim m_c^2$. For realistic
charmonium $v_{c}^{2}\simeq 0.3$, which is comparable with the value of 
$\alpha _{s}(2m_{c}^{2})\simeq 0.3$.  This indicates  that  the colour-octet
mechanism can potentially provide a sufficiently large correction, which  can
be associated with the soft-overlap mechanism (the ultrasoft gluons in the
Coulomb limit). Such  corrections are  sensitive to a long distance
behaviour of the  charmonium  and hadronic wave functions.  This could
lead to a strong violation of the  ratio $Q$ which is expected from the hard
annihilation which depends only from the wave function at the origin
 and therefore  one expects that
\begin{equation}
Q=\frac{\text{Br}[J/\psi \rightarrow p\bar{p}]}{\text{Br}[\psi (2S)\rightarrow
p\bar{p}]}\simeq \frac{\text{Br}[J/\psi \rightarrow e^{+}e^{-}]}{\text{Br}%
[\psi (2S)\rightarrow e^{+}e^{-}]}.
\end{equation}%
However  this relation is satisfied to a very good accuracy  $Q_{p\bar{p}}=0.139$, $Q_{e^{+}e^{-}}=0.133$.  This observation allows one to assume that
the  dominant effects in $p\bar p$ decay  is provided by colour-singlet mechanism,  which  is
proportional to the charmonium wave function at the origin.  

The expression for the decay width reads 
\begin{equation}
\Gamma \lbrack J/\psi \rightarrow p\bar{p}]=\frac{M_{\psi }\beta }{12\pi }%
\left( \left\vert \mathcal{G}_{M}\right\vert ^{2}+\frac{2m_{N}^{2}}{M_{\psi
}^{2}}\left\vert \mathcal{G}_{E}\right\vert ^{2}\right) , 
 \label{width}
\end{equation}%
where we introduced the helicity amplitudes 
\bea
\mathcal{G}_{M}=A_{1}+A_{2}, \quad \mathcal{G}_{E}=A_{1}+\frac{M_{\psi }^{2}}{4m_{N}^{2}}A_{2}.
\eea
Using Eqs (\ref{A1:res}) and (\ref{A2:res}) one finds
\bea
\mathcal{G}_{M}\simeq A^{(0)}_{1}=A_0 I_0, \quad \mathcal{G}_{E}=A_{0}~\frac14 J.
\eea
where we introduced the convenient normalisation factor
\begin{equation}
A_{0}=\frac{f_{\psi }}{~m_{c}^{2}}\frac{~f_{N}^{~2}}{m_{c}^{4}}~(\pi \alpha
_{s})^{3}\frac{10}{81}.  \label{A0}
\end{equation}%

The exact expression for  the  coefficient $\alpha$ describing the angular distribution in  Eq.(\ref{alpha:def}) reads
\begin{equation}
\alpha =\frac{\left\vert \mathcal{G}_{M}\right\vert ^{2}-\frac{4m_{N}^{2}}{%
M_{\psi }^{2}}\left\vert \mathcal{G}_{E}\right\vert ^{2}}{\left\vert 
\mathcal{G}_{M}\right\vert ^{2}+\frac{4m_{N}^{2}}{M_{\psi }^{2}}\left\vert 
\mathcal{G}_{E}\right\vert ^{2}}.  \label{alpha:exp}
\end{equation}
 If   $A_{1}\gg A_{2}$,  then  $A_{2}$ can be neglected  in $\mathcal{G}_{M}$ and $\mathcal{G}_{E}$  and one receives the expression from Eq.(\ref{alfkin}). 
 The nice feature of  this observable is that it does not depend on the normalisation $A_{0}$ which has large uncertainty from the values of $\alpha_{s},\ m_c$ and $f_\psi$. 

 The accurate measurements carried out by BESIII provide  \cite{Ablikim:2012eu}
\begin{equation}
\alpha =0.595\pm 0.012.
\end{equation}%
From the known $\alpha$  one easily gets  the ratio $\left\vert \mathcal{G}_{E}\right\vert /\left\vert \mathcal{G}_{M}\right\vert $ \
\begin{equation}
\left\vert \mathcal{G}_{E}\right\vert /\left\vert \mathcal{G}_{M}\right\vert
=0.832\pm 0.015.
\label{REM}
\end{equation}%
This result  allows one  to conclude  that the effect from the amplitude  $A_{2}$  is not negligible if one wants to accurately get  the value  of $\alpha$.

Below we consider the qualitative analysis of $\alpha$ and  the branching ratio.  
The integrals $I_{0}$ and $J$ are given in Eqs.(\ref{def:I}) and (\ref{J12345}). Their values depend on the models of  LCDAs.

 For the nucleon twist-3 LCDA we will use the model with the truncated  conformal expansion from Ref.\cite{Anikin:2013aka}
\begin{eqnarray}
\varphi _{3}(x_{i}) \simeq 120x_{1}x_{2}x_{3}\left( 1+\varphi _{10}%
\mathcal{P}_{10}(x_{i})+\varphi _{11}\mathcal{P}_{11}(x_{i})\right.
\left. +\varphi _{20}\mathcal{P}_{20}(x_{i})+\varphi _{21}\mathcal{P}%
_{21}(x_{i})+\varphi _{22}\mathcal{P}_{22}(x_{i})\right)
\label{def:phi3} 
\end{eqnarray}%
where 
\begin{equation}
\mathcal{P}_{10}(x_{i})=21(x_{1}-x_{3}),~~\mathcal{P}%
_{11}(x_{i})=7(x_{1}-2x_{2}+x_{3}),  \label{P1i}
\end{equation}%
\begin{align}
&\mathcal{P}_{20}(x_{i}) =\frac{63}{10}\left[
3(x_{1}-x_{3})^{2}-3x_{2}(x_{1}+x_{3})+2x_{2}^{2}\right] ,~
\\
& \mathcal{P}_{21}(x_{i}) =\frac{63}{2}(x_{1}-3x_{2}+x_{3})(x_{1}-x_{3}),
 \\
& \mathcal{P}_{22}(x_{i}) =\frac{9}{5}\left[
x_{1}^{2}+9x_{2}(x_{1}+x_{3})-12x_{1}x_{3}-6x_{2}^{2}+x_{3}^{2}\right] .
  \label{P2i} 
\end{align}%
The coupling $f_{N}$ and the coefficients $\varphi _{ij}$ are multiplicatively
renormalizable and corresponding anomalous dimensions can be found in Refs.\cite{Braun:2008ia, Anikin:2013aka}.

The required twist-4 LCDAs reads%
\begin{equation}
\Phi _{4}(x_{i})=f_{N}~\Phi _{4}^{WW}(x_{i})+\lambda _{1}\bar{\Phi}%
_{4}(x_{i}),  \label{defPhi4}
\end{equation}%
\begin{equation}
\Psi _{4}(x_{i})=f_{N}~\Psi _{4}^{WW}(x_{i})-\lambda _{1}\bar{\Psi}%
_{4}(x_{i}).  \label{def:Psi4}
\end{equation}%
The functions with the  index $WW$ correspond to the Wandzura-Wilczek
contributions, which are defined by  the $\varphi _{3}(x_{i})$.  The explicit
expressions for these functions were obtained in Refs. \cite{Braun:2008ia, Anikin:2013yoa}. 
To our  accuracy they read 
\begin{align}
\Phi _{4}^{WW} (x_i)\simeq &-40\left( 2-\frac{\partial }{\partial x_{3}}%
\right) x_{1}x_{2}x_{3} -20\sum_{k=0}^1\varphi _{1k}\left( 3-\frac{\partial }{\partial x_{3}}\right)
x_{1}x_{2}x_{3}\mathcal{P}_{1k}(x_{i}) \notag \\
&-12\sum_{k=0}^2\varphi _{2k}\left( 4-\frac{\partial }{\partial x_{3}}\right)
x_{1}x_{2}x_{3}\mathcal{P}_{2k}(x_{i}).
\lab{Phi4WW}
\end{align}
The formula for $\Psi _{4}^{WW}(x_{i})$ can be obtained from (\ref{Phi4WW}) by following substitutions in the {\it rhs}: $\partial /\partial x_{3}\to \partial /\partial x_{2}$ and 
$\mathcal{P}_{nk}(1,2,3)\to \mathcal{P}_{nk}(2,1,3)$.
Notice that the differentiations must be computed with the unmodified  expressions of the polynomials $\mathcal{P}_{nk}(x_{i})$  in Eqs.(\ref{P1i}) and (\ref{P2i}) 
and only after that  one can apply the condition $x_{1}+x_{2}+x_{3}=1$.

For the genuine twist-4 functions in Eqs. (\ref{defPhi4}) and (\ref{def:Psi4}) we also use the truncated  conformal expansions
\begin{align}
\bar{\Phi}_{4}(x_{1},x_{2},x_{3})=24x_{1}x_{2}\left( 1+\eta _{10}\mathcal{R%
}_{10}(x_{3},x_{1},x_{2})-\eta _{11}\mathcal{R}_{11}(x_{3},x_{1},x_{2})%
\right) ,
\\
\bar{\Psi}_{4}(x_{1},x\,_{2},x_{3})=24x_{1}x_{3}\left( 1+\eta _{10}\mathcal{R%
}_{10}(x_{2},x_{3},x_{1})+\eta _{11}\mathcal{R}_{11}(x_{2},x_{3},x_{1})%
\right) ,
\end{align}%
where%
\begin{equation}
\mathcal{R}_{10}(x_{1},x_{2},x_{3})=4\left( x_{1}+x_{2}-3/2x_{3}\right) ,~\ 
\mathcal{R}_{11}(x_{1},x_{2},x_{3})=\frac{20}{3}\left( x_{1}-x_{2}+x_{3}/2 \right) .
\end{equation}%
The twist-4 moments  $\lambda _{1}$, $\eta _{10\text{ }}$ and $\eta _{11}$
are  multiplicatively renormalisable, see the details in Refs. \cite{Braun:2008ia, Anikin:2013aka}.

The  four-dimensional  convolution integrals can be easily computed numerically. The explicit results for different convolution integrals are presented  in Appendix \ref{AppC}.

 For our estimates we consider the parameters obtained from the QCD sum
rules \cite{Braun:2000kw}  and from the analysis of the light-cone sum rule for the nucleon electromagnetic form factors  (ABO model) \cite{Anikin:2013aka}. 
We also consider the models with the  parameters  which were recently  obtained in lattice calculations in Ref.\cite{Bali:2019ecy}. 

As a  different  scenario, we  consider for twist-3 DA $\varphi _{3}$
 the model $\varphi ^{II}$ from Ref.\cite{Chernyak:1987nv} (COZ-model). 
 The twist-4 DAs in this case  include the appropriate
WW-terms and we also add the genuine twist-4 moments. Therefore, we denote  this model with the sign plus.
The values of the corresponding parameters are given in the Table~\ref{LCDAmodels}.
 \begin{table}[th]
\centering
\begin{tabular}{| l | l | l | l | l | l | l | l | l | l | } 
\hline
model & $f_{N},\text{GeV}^{2}$ & $\varphi _{10}$ & $\varphi _{11}$ & $\varphi _{20}$ & $\varphi _{21}$ & $\varphi _{22}$ & $\lambda _{1}/f_{N}$
& $\eta _{10}$ & $\eta _{11}$ \\ \hline
ABO & $4.8\times 10^{-3}$ & $0.047$ & $0.047$ & $0.069$ & $-0.024$ & $0.15$ & $%
-6.27$ & $-0.037$ & $0.13$ \\ \hline
Lattice & $3.5\times 10^{-3}$ & $0.051$ & $0.033$ & $0$ & $0$ & $0$ & 
$-12.68$ & $0$ & $0$ \\ \hline
COZ+ & $4.8\times 10^{-3}$ & $0.154$ & $0.182$ & $0.38$ & $0.054$ & 
$-0.146$ & $-6.27$ & $-0.037$ & $0.13$ \\ \hline
\end{tabular}%
\caption{ The LCDA parameters for the different models at  $\mu^{2}=4\text{~GeV}^{2}$ .}  
\label{LCDAmodels}
\end{table}

In our estimates for the branching ratio we always assume  $m_{c}=1.48~$GeV$^{2}$ and  take for the renormalisation scale two values $\mu^{2}=2m_c^2$ and $\mu^{2}=1.5$~GeV$^2$, which correspod to  $\alpha_{s}\simeq 0.30$ and   $\alpha_{s}\simeq 0.35$, respectively. The value of $f_{\psi }$ is fixed by Eq.(\ref{R10BT}).  
Performing the required calculations we obtain the following results.

\emph{ABO-model.}  Let us consider first the ratio (\ref{REM}), which is less sensitive to the value of the factorisation scale $\mu$.  We obtain
\bea
\frac{  \mathcal{G}_E } { \mathcal{G}_M }=\frac{J}{4 I_0}
=0.67 ,\quad (\text{ or }\alpha=0.71).
\label{REMABO}
\eea
This result is valid for all values of $\mu$ as described above.  The obtained  value is about $20$\%  smaller than the experimental  value  in Eq.(\ref{REM}) and can be accepted as a reasonable  leading-order approximation.  The dominant numerical  effect in $ \mathcal{G}_{E}$ is provided by the chiral-odd integral $J_5$ (\ref{J12345}) or more explicitly: by the asymptotic term in WW contribution and by the  genuine twist-4  contribution $\sim \lambda_1\eta_{11}$, see more details in (\ref{J5num}).    Neglecting the contributions of  the integrals  $J_{1,2,3,4}$  and  the momenta $\varphi_{ij}$  in  (\ref{J5num}) one finds
\bea
\left. \frac{J}{4 I_0}\right |_{J_{1,2,3,4}=0,\, \varphi_{ij}=0}=0.23-0.41\frac{\lambda_1\eta_{11}}{f_N}=0.558,
\eea
  which must be compared with the exact value in Eq.(\ref{REMABO}).  This is a good illustration of the important role, which is provided by the  3-quark configuration with the orbital momentum $L_q=1$.  
  The effect of the amplitude  $\mathcal{G}_E$ in the branching ratio is small because it is suppressed by the factor $2m^2_N/M^2_\psi$, see Eq. (\ref{width}).  For the different choices of the factorisation scale $\mu^2=2m_c^2-1.5$~GeV$^2$  we obtain
  \bea
  10^{3}\text{Br}[J/\psi \rightarrow p\bar{p}]\simeq  0.47-1.43,
  \eea 
 while  the experimental branching ratio reads  \cite{Ablikim:2012eu} 
\begin{equation}
10^{3}\text{Br}[J/\psi \rightarrow p\bar{p}]_{\text{exp}}\simeq 2.112\pm 0.004.
\end{equation}
Hence we observe, that  a reliable description of the branching in this model can only be obtained for  the relatively small  values of the renormalisation scale $\mu$.

\emph{Lattice model.} In this case the coupling $ f_{N}$ is smaller, see Table~\ref{LCDAmodels}  and  genuine twist-4 parameters $\eta_{ij}$  are unknown.  If one sets them   
 to zero $\eta _{10}=0$ and $\eta _{11}=0$, one gets very small value for the ratio
 \bea
\frac{  \mathcal{G}_E } { \mathcal{G}_M }=\frac{J}{4 I_0}
=0.10,\quad (\alpha=0.99).
\label{REMLAT}
\eea  
  Because the normalisation constant $f_N$  is also smaller, one also obtains  much  smaller  branching ratio
\begin{equation}
10^{3}\text{Br}[J/\psi \rightarrow p\bar{p}]= 0.08-0.22,
\end{equation}

The value of the ratio $\alpha$ can be easily  improved taking into account  the higher order  twist-4  term with $\eta_{11}$. For instance, taking $\eta_{11}=0.09$
 yields
  \bea
\frac{  \mathcal{G}_E } { \mathcal{G}_M }=\frac{J}{4 I_0}
=0.77,\quad (\alpha=0.64).
\label{REMLATeta11}
\eea  
On the other hand, it is very difficult to improve the value of the branching fraction. Hence, we can conclude that for such a small $f_N$  it is very likely that a large effect from various  corrections or an effect from another decay mechanism is relevant. 

\emph{COZ+ model. } In this model it is very important to take into
account the higher coefficients $\varphi _{2i}$, which provide a strong numerical effect.  
Corresponding LCDA  $\varphi_3$ provides sufficiently large value
for the leading-twist integral $I_{0}$.  This results in the large value of the amplitude $\mathcal{G}_M$ and reduces the value of  the ratio
\bea
\frac{  \mathcal{G}_E } { \mathcal{G}_M }=0.30 ,\quad (\alpha=0.93).
\eea
Taking $\lambda_1=0$ one obtains even smaller value $ \mathcal{G}_E /\mathcal{G}_M =0.20$ . 
At the same time the value of the branching ratio  is large
\footnote{The numerical difference of this result with  Ref.\cite{Chernyak:1987nv} is explained by the different values of $f_\psi$ }
\begin{equation}
10^{3}\text{Br}[J/\psi \rightarrow p\bar{p}] \simeq 4.3-16.1\, .
\end{equation}%
where, remind, we assume the variation of the scale  $\mu^2=1.5-2m_c^2$. 
 It seems, that the leading-twist contribution in this case is overestimated that  strongly reduces the value  $\mathcal{G}_E /  \mathcal{G}_M$.

The considered  set of the LCDA models is not comprehensive. There are many other  models for DA $\varphi_3$,  which were discussed  in the literature, 
see {\it e.g.} Refs. \cite{Bolz:1996sw, King:1986wi, Stefanis:1992nw}.
A more accurate  phenomenological consideration must also include the electromagnetic contribution, which  describes the subprocess $J/\psi\to \gamma^*\to p \bar p$.
 The main conclusion, which follows from the present  calculations, is  a qualitative  estimate of the possible effect  from  the helicity flip amplitudes $\mathcal{G}_E$.
 We find that  the obtained  result for the helicity flip amplitude $\mathcal{G}_E$ is quite sensitive to the genuine twist-4 nucleon DAs and can be quite large comparing 
with  the well known leading-twist contribution in $\mathcal{G}_{M}$.  However the contribution of $\mathcal{G}_E$ to the width is numerically  reduced by the power $m_N^2/M_\psi^2$ in Eq.(\ref{width}) and therefore does not  provide a strong numerical impact on  the  value of the branching fraction.   
Qualitatively this provides a reliable description of the data using the models of DA motivated by the light-cone  QCD sum rules \cite{Anikin:2013aka}.   
Such a  picture suggests that the  twist-four  LCDAs, describing the three quarks with the  orbital angular momentum $L=1$,  are very important  for the description of the 
angular behaviour of the cross section $e^+e^-\to J/\psi\to p\bar p$ . 

\section{Discussion}
\lab{discuss}
The power suppressed amplitude $A_{2}$, which
describes the Pauli ($\sigma ^{\mu \nu }$) vertex  in the    $J/\psi
\rightarrow p\bar{p}$  decay amplitude,  is calculated.  It is shown that at least in the leading-order this amplitude is described  within
the standard collinear QCD framework which is based on the factorisation of the
hard  and soft processes.  The obtained result is used  for a qualitative
phenomenological analysis of the angular distribution of the cross section $e^{+}e^{-}$ $\rightarrow J/\psi \rightarrow p\bar{p}$ and for the decay
width. 

We use different models of nucleon DAs  in order to  compare obtained results with the existing data. The best description is obtained  with the set of DAs obtained from  light-cone QCD sum rules \cite{Anikin:2013aka}.  In this case we find a reliable estimate for the ratio $\mathcal{G}_{E}/\mathcal{G}_{M}$, which describes the angular behaviour of the cross section.   This leads to the conclusion  that the  Fock component of the nucleon wave
function  associated with the  three quarks in $P$-wave state  provides an important effect in the description of  the $J/\psi$ decay.  This allows one to conclude that  quarkonia  decays  into baryon-antibaryon  provide us  an interesting and important  insight about  baryon wave functions. 
 
  The obtained description of the decay width  is worse, one obtains acceptable estimate only taking the relatively low value of the  renormalisation  scale  $\sim 1.5$~GeV$^2$.  This observation opens  questions about the size of the next-to-leading and power corrections to the amplitude $A_1$.  For better understanding of the decay mechanism such corrections must be computed.   

The lattice data \cite{Bali:2019ecy}  suggests  relatively small value of the non-perturbative normalisation constant $f_N$, which makes the description of the width very problematical, because the amplitude $\mathcal{G}_{M}$ turns out to be very small.  Therefore if such  value of the  $f_N$ is correct, one must expect a large contribution from other decay mechanism.   The COZ model \cite{Chernyak:1984bm}  works quite differently, it provides a  large value of the width but on the other hand gives  very small value of the ratio  $\mathcal{G}_{E}/\mathcal{G}_{M}$. Most likely,  this indicates that  the value of the  amplitude $A_1$ in this case is somewhat overestimated.

\section*{\Huge Appendix}
\appendix
\numberwithin{equation}{section}
\setcounter{equation}{0}
\section{ Long distance matrix elements }
\lab{AppA}

Here we provide a brief  summary of the required nonperturbative matrix elements and LCDAs.
For the heavy quark sector we only need the  NRQCD matrix element%
\begin{equation}
\left\langle 0\right\vert \chi _{\omega }^{\dag }(0)\gamma^{\mu}\psi _{\omega }(0)\left\vert P\right\rangle
 =\epsilon _{\psi }^{\mu }~f_{\psi}. 
  \label{def:fpsi}
\end{equation}%
The
operator in (\ref{def:fpsi}) is constructed from the quark $\psi _{\omega }$ and antiquark $%
\chi _{\omega }^{\dag }$ four-component spinor fields satisfying $\NEG{\omega}\psi _{\omega }=\psi _{\omega }$, $\NEG{\omega}\chi _{\omega }=-\chi
_{\omega }$.
  The coupling $f_\psi$  is related with the radial wave function at
the origin
\begin{equation}
f_{\psi }=\sqrt{2M_{J/\psi }}\sqrt{\frac{3}{2\pi }}~R_{10}(0).
\end{equation}%
The value $R_{10}(0)$  is well known from various potential models, for
instance for the Buchmuller-Tye potential \cite{Eichten:1995ch}
\begin{equation}
\left\vert R_{10}(0)\right\vert ^{2}\simeq 0.81\text{GeV}^{3}.
\label{R10BT}
\end{equation}

One can also estimate this coupling from $J/\psi \rightarrow e^{+}e^{-}$
decay using the well known formula for the leptonic width
\begin{equation}
\Gamma \lbrack J/\psi \rightarrow e^{+}e^{-}]=\frac{16}{9}\frac{\alpha
_{em}^{2}}{M_{\psi }^{2}}\left\vert R_{10}(0)\right\vert ^{2}\left( 1-\frac{%
16}{3}\frac{\alpha _{s}}{\pi }\right) .
\end{equation}
This gives $(~$Br$[J/\psi \rightarrow e^{+}e^{-}]=5.97\%,~\alpha
_{s}=0.3,~\alpha _{em}=1/130)$%
\begin{equation}
\left\vert R_{10}(0)\right\vert ^{2}\simeq 0.76~\text{GeV}^{3},
\end{equation}%
which is quite close to the value (\ref{R10BT}).

The nucleon matrix elements are more complicated. In the definitions given below we
use kinematics and notations introduced in the Section 2.  For simplicity,
in this Appendix we consider the matrix elements only for the nucleon state and
we also imply the light-cone gauge
\begin{equation}
n\cdot A^{(\bar{n})}(x)=0,
\end{equation}%
in order to simplify the formulas.

The twist-3 DAs are defined as ($i,j,k$ are the colour indices)
\begin{align}
& \left\langle 0\left\vert  \varepsilon ^{ijk}u_{\alpha }^{i}(z_{1-})u_{\beta
}^{j}(z_{2-})d_{\sigma }^{k}(z_{2-})\right\vert k\right\rangle _{\text{tw3}}
=\frac{1}{4}~\left[ \NEG{k}~C\right] _{\alpha \beta }\left[ \gamma _{5}
N_{\bar{n}}\right] _{\sigma }~\text{FT}\left[ V_{1}(y_{i})\right]
\notag \\
 &\mskip 100mu +\frac{1}{4}\left[ \NEG{k}\gamma _{5}C\right] _{\alpha \beta }\left[ N_{\bar{n}}\right]
_{\sigma }~\text{FT}\left[ A_{1}(y_{i})\right] 
+\frac{1}{4}~\left[ i\sigma _{\bot k}C\right] _{\alpha \beta }\left[
\gamma ^{\bot }\gamma _{5}N_{\bar{n}}\right] _{\sigma }\text{FT}\left[
T_{1}(y_{i})\right] ,
\label{tw3me}
\end{align}%
where%
\begin{equation}
\text{FT}\left[ F(y_{i})\right] =\int Dy_{i}~e^{-iy_{1}k_{-}^{\prime
}z_{1+}/2-iy_{2}k_{-}^{\prime }z_{2+}/2-iy_{3}k_{-}^{\prime
}z_{3+}/2}F(y_{1,}y_{2},y_{3}),
\end{equation}%
with 
\begin{equation}
Dy_{i}=dy_{1}dy_{2}dy_{3}\delta (1-y_{1}-y_{2}-y_{3}).
\end{equation}%
We also explicitly write  the large component of the nucleon spinor%
\begin{equation}
N_{\bar{n}}=\frac{\Dsl{\bar{n}}\Dsl{n}}{4}N(k).
\end{equation}%
Three DAs $V_{1}$, $A_{1}$ and $T_{1}$ can be combined into the one twist-3
DA $\varphi _{3}$ as%
\begin{equation}
V_{1}(x_{1},x_{2},x_{3})=f_{N}\frac{1}{2}\left[ \varphi
_{3}(x_{1},x_{2},x_{3})+\varphi _{3}(x_{2},x_{1},x_{3})\right]
\end{equation}%
\begin{equation}
A_{1}(x_{1},x_{2},x_{3})=f_{N}\frac{1}{2}\left[ \varphi
_{3}(x_{2},x_{1},x_{3})-\varphi _{3}(x_{2},x_{1},x_{3})\right]
\end{equation}%
\begin{equation}
T_{1}(x_{1},x_{2},x_{3})=f_{N}\frac{1}{2}\left[ \varphi
_{3}(x_{1},x_{3},x_{2})+\varphi _{3}(x_{2},x_{3},x_{1})\right] .
\lab{T1[phi3]}
\end{equation}

The twist-4 LCDAs  are defined as 
\begin{align}
4\left\langle 0\left\vert \varepsilon ^{ijk}u_{\alpha }^{i}(z_{1-})u_{\beta
}^{j}(z_{2-})~d_{\sigma }^{k}(z_{3-})\right\vert k\right\rangle _{\text{tw4}}
=m_{N}\left[ C\right] _{\alpha \beta }~\left[ \gamma _{5}N_{\bar{n}}\right]
_{\sigma }~\text{FT}\left[ S_{1}(x_{i})\right] 
\notag \\
+m_{N}\left[ \gamma _{5}C\right] _{\alpha \beta }~\left[ N_{\bar{n}}\right] _{\sigma }~\text{FT}\left[
P_{1}(x_{i})\right]
+\frac{1}{4}m_{N}\left[ \nbs C\right] _{\alpha \beta }~\left[ \gamma _{5}%
\NEG{n}N_{\bar{n}}\right] _{\sigma }~\text{FT}\left[ V_{2}(x_{i})\right]
\notag\\
+\frac{1}{2}m_{N}\left[ \gamma _{\bot }C\right] _{\alpha \beta }~\left[
\gamma ^{\bot }\gamma _{5}N_{\bar{n}}\right] _{\sigma }~\text{FT}\left[
V_{3}(x_{i})\right]
+\frac{1}{4}m_{N}\left[ \nbs \gamma _{5}C\right] _{\alpha \beta }~\left[ 
\NEG{n}N_{\bar{n}}\right] _{\sigma }~\text{FT}\left[ A_{2}(x_{i})\right]
\notag \\
+\frac{1}{2}m_{N}\left[ \gamma _{\bot }\gamma _{5}C\right] _{\alpha \beta }~%
\left[ \gamma ^{\bot }N_{\bar{n}}\right] _{\sigma }~\text{FT}\left[
A_{3}(x_{i})\right]
+\frac{1}{4}m_{N}\left[ \sigma _{\bot -}C\right] _{\alpha \beta }~%
\left[ \gamma ^{\bot }\gamma _{5}\NEG{n}N_{\bar{n}}\right] _{\sigma }~\text{%
FT}\left[ T_{2}(x_{i})\right]
\notag \\
+\frac{1}{2}m_{N}\left[ \sigma _{-+}C\right] _{\alpha \beta }~\left[
\gamma _{5}N_{\bar{n}}\right] _{\sigma }~\text{FT}\left[ T_{3}(x_{i})\right]
+\frac{1}{2}m_{N}\left[ \sigma _{\bot \bot ^{\prime }}C\right] _{\alpha
\beta }~\left[ \sigma ^{\bot \bot ^{\prime }}\gamma _{5}N_{\bar{n}}\right]
_{\sigma }~\text{FT}\left[ T_{7}(x_{i})\right] , 
 \label{LCmetw4}
\end{align}
where
\begin{equation}
\sigma _{-+}=\sigma _{\mu \nu }\bar{n}^{\mu }n^{\nu }.
\end{equation}%
These LCDAs  can be written in terms of  three twist-4 LCDAs in the following  \footnote{We use the definitions of nucleon DAs from Ref.\cite{Anikin:2013aka}. Let us notice that definitions of the twist-4 DAs  $\Psi_4$, $\Phi_4$ and $\Xi_4$ in Ref.\cite{Braun:2000kw}  are different. }
\begin{equation}
V_{2}(x_{1},x_{2},x_{3})=\frac{1}{4}\left[ \Phi _{4}(x_{1},x_{2},x_{3})+\Phi
_{4}(x_{2},x_{1},x_{3})\right] ,
\end{equation}%
\begin{equation}
V_{3}(x_{1},x_{2},x_{3})=\frac{1}{4}\left[ \Psi _{4}(x_{1},x_{2},x_{3})+\Psi
_{4}(x_{2},x_{1},x_{3})\right] ,
\end{equation}%
\begin{equation}
A_{2}(x_{1},x_{2},x_{3})=\frac{1}{4}\left[ \Phi _{4}(x_{2},x_{1},x_{3})-\Phi
_{4}(x_{1},x_{2},x_{3})\right] ,
\end{equation}%
\begin{equation}
A_{3}(x_{1},x_{2},x_{3})=\frac{1}{4}\left[ \Psi _{4}(x_{2},x_{1},x_{3})-\Psi
_{4}(x_{1},x_{2},x_{3})\right] ,
\end{equation}%
\begin{equation}
T_{3}(x_{1},x_{2},x_{3})=\frac{1}{8}\left[\frac{\lambda_2}3 \Xi _{4}(x_{1},x_{2},x_{3})+\Psi
_{4}(x_{3},x_{1},x_{2})+\Phi _{4}(x_{2},x_{3},x_{1})\right]
+(x_{1}\leftrightarrow x_{2}),
\end{equation}%
\begin{equation}
S_{1}(x_{1},x_{2},x_{3})=\frac{1}{8}\left[\frac{\lambda_2}3  \Xi _{4}(x_{1},x_{2},x_{3})+\Psi
_{4}(x_{3},x_{1},x_{2})+\Phi _{4}(x_{2},x_{3},x_{1})\right]
-(x_{1}\leftrightarrow x_{2}),
\end{equation}%
\begin{equation}
T_{7}(x_{1},x_{2},x_{3})=\frac{1}{8}\left[ -\frac{\lambda_2}3 \Xi _{4}(x_{1},x_{2},x_{3})+\Psi
_{4}(x_{3},x_{1},x_{2})+\Phi _{4}(x_{2},x_{3},x_{1})\right]
+(x_{1}\leftrightarrow x_{2}),
\end{equation}%
\begin{equation}
P_{1}(x_{1},x_{2},x_{3})=-\frac{1}{8}\left[ -\frac{\lambda_2}3 \Xi
_{4}(x_{1},x_{2},x_{3})+\Psi _{4}(x_{3},x_{1},x_{2})+\Phi
_{4}(x_{2},x_{3},x_{1})\right] -(x_{1}\leftrightarrow x_{2}).
\end{equation}

In our calculation we use the matrix elements of twist-4 operators
constructed from the large collinear components $\chi _{\bar{n}}$ (\ref{chi})
 and their derivative $\partial _{\bot }\chi _{\bar{n}}$,  see Eq.(\ref{OV1}).
  In order to find expressions for these matrix elements we need to
consider off light-cone correlators.  The chiral even correlators have
already been considered in Ref.\cite{Anikin:2013aka}. Consider, for simplicity,  the
vector projection. The corresponding correlator reads  
\begin{align}
-\left\langle 0\left\vert \varepsilon ^{ijk}u^{i}(z_{1})C\gamma ^{\alpha
}u^{j}(z_{2})d_{\sigma }^{k}(z_{3})\right\vert k\right\rangle =k^{\alpha } 
\left[ \gamma _{5}N\right] _{\sigma }\text{FT}\left[ V_{1}\right] 
+m_{N}\left[ \gamma ^{\alpha }\gamma _{5}N\right] _{\sigma }\text{FT}\left[
V_{3}\right]
\notag \\
+m_{N}ik^{\alpha }\left( z_{1\beta }~\text{FT}\left[ \mathcal{V}_{1}\right]
+z_{2\beta }~\text{FT}\left[ \mathcal{V}_{2}\right] +z_{3\beta }~\text{FT}%
\left[ \mathcal{V}_{3}\right] \right) \left[ \gamma ^{\beta }\gamma _{5}N%
\right] _{\sigma }.
\lab{Vcorr}
\end{align}
By calligraphic letters we denote the auxiliary LCDAs, which can be rewritten in
terms of defined above in Eq.(\ref{LCmetw4})  twist-4 LCDAs. The explicit expressions will be given
below. Performing expansion of the operator in the \textit{lhs} (\ref{Vcorr}) according to
formulas (\ref{LCexp}),  expanding on the \textit{rhs} the coordinates $z_{i}\simeq (z_{i}\bar{n}%
)n/2+z_{i\bot }$ in $z_{i\bot }$ and comparing the linear in $z_{i\bot }$ contributions one
finds ( in this section we denote $\xi _{\bar{n}}(x)\equiv \xi (x)$ in order
to simplify notations)%
\begin{equation}
~\left\langle 0\right\vert \varepsilon ^{ijk}\left[ i\partial _{\bot \alpha
}\xi ^{i}(z_{1-})\right] C\NEG{n}\xi ^{j}(z_{2-})\xi _{\sigma
}^{k}(z_{3-})\left\vert k\right\rangle =k_{+}m_{N}~\left[ \gamma _{\bot\alpha
}\gamma _{5}N_{\bar n}\right] _{\sigma }\text{FT}\left[ \mathcal{V}_{1}%
\right] ,  \label{meV1cal}
\end{equation}%
\begin{equation}
~\left\langle 0\right\vert \varepsilon ^{ijk}\xi ^{i}(z_{1-})C\NEG{n}\left[
i\partial _{\bot \alpha }\xi ^{j}(z_{2-})\right] \xi _{\sigma
}^{k}(z_{3-})\left\vert k\right\rangle =k_{+}m_{N}~\left[ \gamma _{\bot\alpha
}\gamma _{5}N_{\bar n} \right] _{\sigma }\text{FT}\left[ \mathcal{V}_{2}%
\right] .
\end{equation}%
For the axial projection one finds%
\begin{align}
-\left\langle 0\left\vert \varepsilon ^{ijk}u^{i}(z_{1})C\gamma ^{\alpha
}\gamma _{5}u^{j}(z_{2})d_{\sigma }^{k}(z_{3})\right\vert k\right\rangle
=k^{\alpha }\left[ N\right] _{\sigma }\text{FT}\left[ A_{1}\right] +m_{N}%
\left[ \gamma ^{\alpha }N\right] _{\sigma }\text{FT}\left[ A_{3}\right]
\notag \\
+m_{N}ik^{\alpha }\left( z_{1\beta }~\text{FT}\left[ \mathcal{A}_{1}\right]
+z_{2\beta }~\text{FT}\left[ \mathcal{A}_{2}\right] +z_{3\beta }~\text{FT}%
\left[ \mathcal{A}_{3}\right] \right) \left[ \gamma ^{\beta }N\right]
_{\sigma },
\end{align}%
The expansion around the light-cone direction gives%
\begin{equation}
~\left\langle 0\right\vert \varepsilon ^{ijk}\left[ i\partial _{\bot \alpha
}\xi ^{i}(z_{1-})\right] C\NEG{n}\gamma _{5}\xi ^{j}(z_{2-})\xi _{\sigma
}^{k}(z_{3-})\left\vert k\right\rangle =k_{+}m_{N}~\left[ \gamma _{\bot\alpha
} N_{\bar n}\right] _{\sigma }\text{FT}\left[ \mathcal{A}_{1}\right] ,
\end{equation}%
\begin{equation}
~\left\langle 0\right\vert \varepsilon ^{ijk}\xi ^{i}(z_{1-})C\NEG{n}\gamma
_{5}\left[ i\partial _{\bot \alpha }\xi ^{j}(z_{2-})\right] \xi _{\sigma
}^{k}(z_{3-})\left\vert k\right\rangle =k_{+}m_{N}~\left[ \gamma _{\bot\alpha
} N_{\bar n}\right] _{\sigma }\text{FT}\left[ \mathcal{A}_{2}\right] .
\end{equation}

We also need  to consider the  chiral-odd correlator%
\begin{align}
&-\left\langle 0\right\vert \varepsilon ^{ijk}u_{\alpha }^{i}(z_1)C\sigma ^{\mu
\nu }u_{\beta }^{j}(z_2)d_{\sigma }^{k}(z_3)\left\vert k\right\rangle
=~ip^{\nu }\left[ \gamma ^{\mu }\gamma _{5}N\right] \text{FT}\left[ T_{1}\right]
 +~\frac{1}{2}m_{N}~\left[ \sigma ^{\mu \nu }\gamma _{5}N\right] 
\text{FT}\left[ T_{7}\right]
\notag\\   &  \mskip 50 mu
+\left( z_1-z_3\right) ^{\mu }p^{\nu }~m_{N}\left[ \gamma _{5}N\right] \text{FT%
}\left[ \mathcal{T}_{21}\right] 
+~\left( z_2-z_3\right) ^{\mu }p^{\nu }~m_{N}\left[ \gamma _{5}N\right] \text{%
FT}\left[ \mathcal{T}_{22}\right]
\notag \\
&+~m_{N}~ip^{\nu }(z_1-z_3)_{\beta }\left[ \sigma ^{\mu \beta }\gamma _{5}N\right]
\text{FT}\left[ \mathcal{T}_{41}\right]
+~m_{N}~ip^{\nu }(z_2-z_3)_{\beta }\left[ \sigma ^{\mu \beta }\gamma _{5}N\right]
\text{FT}\left[ \mathcal{T}_{42}\right] 
\notag \\ & \mskip300mu
-(\mu \leftrightarrow \nu ).
\end{align}%
This equation yields%
\begin{align}
&\left\langle 0\right\vert \varepsilon ^{ijk}\left[ i\partial _{\bot
}^{\alpha }\xi ^{i}(z_{1-})\right] C\NEG{n}\gamma _{\bot }^{\beta }\xi
^{j}(z_{2-})\xi _{\sigma }^{k}(z_{3-})\left\vert k\right\rangle  =g_{\bot
}^{\alpha \beta }m_{N}k_{+}\left[ \gamma _{5}N_{\bar n}\right] _{\sigma }\text{FT}%
\left[ \mathcal{T}_{21}\right]
\notag \\ & \mskip 400 mu
 -m_{N}k_{+}\left[ i\sigma _{\bot \bot }^{\alpha \beta }\gamma _{5}N_{\bar n}\right]
_{\sigma }\text{FT}\left[ \mathcal{T}_{41}\right] ,
\end{align}
\begin{align}
&\left\langle 0\right\vert \varepsilon ^{ijk}\xi ^{i}(z_{1-})C\NEG{n}\gamma
_{\bot }^{\beta }\left[ i\partial _{\bot }^{\alpha }\xi ^{j}(z_{2-})\right]
\xi _{\sigma }^{k}(z_{3-})\left\vert k\right\rangle  =g_{\bot }^{\alpha
\beta }m_{N}k_{+}\left[ \gamma _{5}N_{\bar n}\right] _{\sigma }\text{FT}\left[ 
\mathcal{T}_{22}\right]
\notag \\ & \mskip 400 mu
 -m_{N}k_{+}\left[ i\sigma _{\bot \bot }^{\alpha \beta }\gamma _{5}N_{\bar n}\right]
_{\sigma }\text{FT}\left[ \mathcal{T}_{42}\right] .
\end{align}

The  LCDAs which are denoted by calligraphic letters  can be rewritten in
terms of the light-cone LCDAs which are defined by  the light-cone matrix element (\ref{LCmetw4}).  
For the LCDAs $\mathcal{V}_{1,2}$ and $\mathcal{A}_{1,2}$ such
expressions are already derived in Ref.\cite{Anikin:2013aka}. We also 
recalculated these relations and  find the same  expressions.  They read (we assume $f(x_i)\equiv f(x_1,x_2,x_3) $)
\begin{equation}
\mathcal{V}_{1}(x_{i})+\mathcal{V}_{3}(x_{i})+\mathcal{V}_{3}(x_{i})=0,
\end{equation}
\begin{equation}
4\mathcal{V}_{k}(x_{i})=x_{3}V_{2}(x_{i})+(-1)^{k}\left\{
(x_{1}-x_{2})V_{3}(x_{i})-x_{3}A_{2}(x_{i})+\bar{x}_{3}A_{3}(x_{i})\right\} .
\label{Vical}
\end{equation}%
\begin{equation}
\mathcal{A}_{1}(x_{i})+\mathcal{A}_{3}(x_{i})+\mathcal{A}_{3}(x_{i})=0,
\end{equation}%
\begin{equation}
4\mathcal{A}_{k}(x_{i})=-x_{3}A_{2}(x_{i})+(-1)^{k}\left\{
(x_{1}-x_{2})A_{3}(x_{i})+x_{3}V_{2}(x_{i})+\bar{x}_{3}V_{3}(x_{i})\right\} .
\end{equation}%
Notice that%
\begin{eqnarray}
\mathcal{V}_{2}(x_{1},x_{2},x_{3}) =\mathcal{V}_{1}(x_{2},x_{1},x_{3}), 
~\ \mathcal{A}_{2}(x_{1},x_{2},x_{3}) =-~\mathcal{A}_{1}(x_{2},x_{1},x_{3}),
\lab{VA12tw4}
\end{eqnarray}%
which follows from 
\begin{equation}
V_{i}(2,1,3)=V_{i}(1,2,3),~ A_{i}(2,1,3)=-A_{i}(1,2,3).
\lab{VA12}
\end{equation}

The similar relations  for the  chiral-odd LCDAs $\mathcal{T}_{ij}$ have not yet been considered. 
Our calculations yield (see the details below)
\begin{eqnarray}
\mathcal{T}_{21}(x_{i})-\mathcal{T}_{41}(x_{i}) &=&
\frac{x_{1}}{2}\left(T_{3}(x_{i})+T_{7}(x_{i})+S_{1}(x_{i})-P_{1}(x_{i})\right) 
\\  
{}&=&\frac{x_{1}}{2}~\left[ V_{3}(3,1,2)-A_{3}(3,1,2)+V_{2}(2,3,1)-A_{2}(2,3,1)\right] ,
\label{T2mT4}
\\
\mathcal{T}_{22}(x_{i})-\mathcal{T}_{42}(x_{i})&=&\frac{x_{2}}{2}\left(
T_{3} (x_{i})+T_{7} (x_{i})-S_{1} (x_{i})+P_{1} (x_{i})\right)
\\
&=&\frac{x_{2}}{2}~\left[ V_{3}(3,2,1)-A_{3}(3,2,1)+V_{2}(1,3,2)-A_{2}(1,3,2)\right] ,
\end{eqnarray}%
\begin{equation}
\mathcal{T}_{41}(x_{i})+\mathcal{T}_{21}(x_{i})=\frac{x_{1}}{2}\left(
T_{3}(x_{i})-T_{7}(x_{i})+P_{1}(x_{i})+S_{1}(x_{i})\right)=\frac{x_{1}}{2}\frac{\lambda_2}{6}\Xi _{4}(x_{i}),
\end{equation}%
\begin{equation}
\mathcal{T}_{42}(x_i)+\mathcal{T}_{22}(x_i)=\frac{x_{2}}{2}\left(
T_{3}(x_i)-T_{7}(x_i)-S_{1}(x_i)-P_{1}(x_i)\right) =\frac{x_{2}}{2}\frac{\lambda_2}{6}\Xi _{4}(2,1,3),
\label{T42mnT22}
\end{equation}
where it was used that
\begin{equation}
T_{i}(2,1,3)=T_{i}(1,2,3),~
 S_{1}(2,1,3)=-S_{1}(1,2,3),~P_{1}(2,1,3)=-~P_{1}(1,2,3).
\end{equation}%

 Consider the derivation of Eqs. (\ref{T2mT4})-(\ref{T42mnT22}). 
 Let us introduce two twist-4 light-cone operators  defined as 
\begin{equation}
O_{1}=\left[ \frac{\ns\nbs}{4}u(x_{-})\right] C\sigma ^{\mu \nu }
\frac{\nbs \ns}{4}u(y_{-})\left[ \frac{\nbs \ns }{4}d(z_{-})\right] _{\sigma },
\label{defO1}
\end{equation}%
\begin{equation}
O_{2}=\left[ \frac{\nbs\ns }{4}u(x_{-})\right] C\sigma ^{\mu \nu }
\frac{\ns\nbs }{4}u(y_{-})\left[ \frac{\nbs\ns}{4}d(z_{-})\right] _{\sigma },
\label{defO2}
\end{equation}%
where  the projectors $\Dsl{\bar{n}}\Dsl{n}/4$ and $\Dsl{n}\Dsl{\bar{n}}/4$ are used in order to decompose 
  collinear fields into large and small components, respectively
\begin{equation}
\frac{\Dsl{\bar{n}}\Dsl{n} }{4}u=\xi ,~\frac{\Dsl{n}\Dsl{\bar{n}} }{4}u(x_{-})=\eta .
\end{equation}
Using Eq.(\ref{etaEOM}) we rewrite the first operator as 
\begin{equation}
O_{1}=-T_{1}^{\mu \nu \alpha \lambda }\left[ \left( in\partial \right)
^{-1}\partial _{\bot \alpha }\xi (x_{-})\right] C\sigma ^{+\lambda }\xi
(y_{-})\left[ \xi (z_{-})\right] _{\sigma }.
\end{equation}
where 
\begin{equation}
T_{1}^{\mu \nu \alpha \lambda }=\left\{ g_{\bot }^{\alpha \nu }g_{\bot
}^{\lambda \mu }~+\frac{1}{2}~\bar{n}^{\mu }n^{\nu }g_{\bot }^{\lambda
\alpha }~-(\mu \leftrightarrow \nu )\right\} .
\end{equation}%
Taking the matrix element with the help of Eq.(\ref{meV1cal})  one obtains%
\begin{equation}
-\left\langle 0\right\vert O_{1}\left\vert k\right\rangle =~i\bar{n}^{\mu
}n^{\nu }\frac{m_{N}}{2}\left[ \gamma _{5}N_{\bar{n}}\right] _{\sigma }~%
\text{FT}\left[ \frac{1}{x_{1}}\mathcal{T}_{21}\right]
 -m_{N}\frac{1}{2}\left[
\sigma _{\bot \bot }^{\mu \nu }\gamma _{5}N_{\bar{n}}\right] _{\sigma }~%
\text{FT}\left[ \frac{1}{x_{1}}\mathcal{T}_{41}\right] -(\mu \leftrightarrow
\nu ).  \label{O1me}
\end{equation}

On the other hand rewriting the operator (\ref{defO1}) with the basic Dirac structures one finds
\begin{align}
&O_1
 =\frac{i}{4}\left( \bar{n}^{\nu }n^{\mu }-\bar{n}^{\mu }n^{\nu }\right)
~u(x_{-})Cu(y_{-})\left[ \frac{\nbs\ns }{4}d(z_{-})\right] _{\sigma } 
\notag \\
& +\frac{1}{2}\varepsilon _{\mu \nu +- }~u(x_{-})C\gamma _{5}u(y_{-})%
\left[ \frac{\nbs\ns}{4}d(z_{-})\right] _{\sigma }
+\frac{1}{2}u(x_{-})C\sigma _{\bot \bot }^{\mu \nu }u(y_{-})\left[ \frac{\nbs\ns}{4}d(z_{-})\right] _{\sigma } 
\notag \\
& +\frac{1}{8}\left( \bar{n}^{\nu }n^{\mu }-\bar{n}^{\mu }n^{\nu }\right)
u(x_{-})C\sigma ^{-+}u(y_{-})\left[ \frac{\nbs\ns }{4}d(z_{-})\right]
_{\sigma }.
\end{align}%
Taking the matrix element in the \textit{rhs} of this equation with the
help of Eq.(\ref{LCmetw4})  and  comparing with the Eq.(\ref{O1me}) one
obtains (for simplicity it is used  $f(x_i)\equiv f$)%
\begin{equation}
\mathcal{T}_{21}-\mathcal{T}_{41}=\frac{x_{1}}{2}\left(
T_{3}+T_{7}+S_{1}-P_{1}\right) , ~~
\mathcal{T}_{21}+\mathcal{T}_{41}=\frac{x_{1}}{2}\left(
T_{3}-T_{7}+P_{1}+S_{1}\right) .
\end{equation}%
The similar consideration for the operator $O_{2}$ in Eq.(\ref{defO2}) gives%
\begin{equation}
\mathcal{T}_{22}-\mathcal{T}_{42}=\frac{x_{2}}{2}\left(
T_{3}+T_{7}-S_{1}+P_{1}\right) ,~~
\mathcal{T}_{42}+\mathcal{T}_{22}=\frac{x_{2}}{2}\left(
T_{3}-T_{7}-S_{1}-P_{1}\right) .
\end{equation}

\section{ The cancellation of the ultrasoft gluon contributions}
\lab{AppB}
Here we briefly discuss the ultrasoft gluon limit.  The contribution of the
sum of diagrams as  in Fig. \ref{hard_diagram} can be written as 
\begin{equation}
iM=\frac{f_{\psi }}{m_{Q}^{2}}\frac{f_{N}\lambda _{1}}{m_{Q}^{4}}\frac{m_{N}%
}{m_{Q}}~J,~  \label{M=J}
\end{equation}%
where the dimensionless collinear convolution integral can be schematically
written as 
\begin{equation}
J=m_{Q}^{7}\int Dx_{i}\int Dy_{i}~\Delta _{g1}\Delta _{g2}\Delta _{g3}~
\hat T_{3g\to p\bar{p}}(x_{i},y_{i})\frac{1}{4}\text{Tr}\left[ (1-\NEG%
{\omega})\NEG{\epsilon}_{\psi }D(k_{i}^{\prime },k_{j})\right] .  \label{J}
\end{equation}%
For simplicity, we do not show  the various indices. Notice also that the
colour factors for all diagrams are the same. In Eq.(\ref{J}) we have
three gluon propagators
\begin{equation}
\Delta _{gi}=\frac{(-i)}{(k_{i}+k_{i}^{\prime })^{2}}\simeq \frac{(-i)}{2(kk^{\prime })}\frac{1}{x_{i}y_{i}}.
\end{equation}%
The function $\hat T_{3g\to p\bar{p}}$ describes  the contribution from  the
light-quark vertices and from the projections  of  the nucleon matrix elements. The
function $D(k_{i}^{\prime },k_{j})$ describes the heavy quark lines with
the quark-gluon vertices.

The scaling behaviour of the contribution in Eq.(\ref{M=J})  is given by the following  factors
\begin{equation}
\frac{f_{\psi }}{m_{Q}^{2}}\sim v^{3},~~\frac{f_{N}~\lambda _{1}}{m_{Q}^{4}}%
\frac{m_{N}}{m_{Q}}\sim \left( \frac{\Lambda }{m_{Q}}\right) ^{5}\sim
\lambda ^{10}.
\end{equation}
The collinear integral  is defined to be of order one: $J\sim $ $v^{0}$. 
 Therefore the  ultrasoft  region in $J$  must  give the contribution  of order one.

Consider the ultrasoft gluon limit 
\begin{equation}
p_{g}=k_{1}+k_{1}^{\prime }\sim m_{Q}v^{2},
\end{equation}%
that gives the counting for the small momentum fractions $x_{1}\sim y_{1}\sim v^{2}$. Such limit corresponds to the
contribution from the endpoint region%
\begin{align}
&J_{us} \sim \int_{0}^{\eta }dx_{1}\int_{0}^{\eta }dy_{1}\frac{1}{x_{1}y_{1}}%
\int_{0}^{1}dx_{2}\int_{0}^{1}dy_{2}~~\Delta _{g2}\Delta _{g3} 
\notag \\ & \mskip 200mu
\times \hat T_{3g\to p\bar{p}}(x_{i},y_{i})\frac{1}{4}\text{Tr}\left[
(1-\NEG{\omega})\NEG{\epsilon}_{\psi }D^{\mu _{1}\mu _{2}\mu
_{3}}(k_{i}^{\prime },k_{j})\right] ,  \label{Jsoft}
\end{align}%
where the cut-off $\eta $ can be understood as a factorisation scale separating the hard and ultrasoft domains. Our task
is to estimate the scale behaviour of $J_{us}$. For that we need to expand
the integrand with respect to small fractions $x_{1}\sim y_{1}\sim v^{2}$.

The light-quark part $\hat T_{3g\to p\bar{p}}(x_{i},y_{i})$ includes the
 twist-4 nucleon LCDAs $\mathcal{V}_{i}(x_{i}),~\mathcal{A}_{i}(x_{i})$ and 
$\mathcal{T}_{ij}(x_{i})$ and twist-3 $\varphi _{3}(y_{i})$. These functions must be
also expanded with respect to the small fractions. We assume that the expansions of the
LCDAs can provide only the positive powers of the small fractions. Therefore, it
is quite reasonable here to consider  only the asymptotic terms, which gives
the contributions with the minimal powers of all fractions. Since one of our LCDAs is of 
twist-3, we immediately find that 
\begin{equation}
\hat T_{3g\to p\bar{p}}\simeq \varphi _{3}^{as}(y_{i})
\hat T^{\text{tw4}}_{3g\to p\bar{p}} (x_{2}) \sim y_{1}y_{2}y_{3} \hat T^{\text{tw4}}_{3g\to p\bar{p}} (x_{2}),
\end{equation}%
where we assume that  $x_{3}\simeq 1-x_{2}$.  The asymptotic expressions
for the twist-4 LCDAs  can be easily obtained from the formulas in Appendix~\ref{AppA}. One finds 
\begin{equation}
\mathcal{V}_{i}(x_{i})\sim x_{1}x_{2}x_{3},~\mathcal{A}_{i}(x_{i})\sim
x_{1}x_{2}x_{3},~\mathcal{T}_{ij}(x_{i})\sim x_{1}x_{2}x_{3}.  \label{asXi}
\end{equation}%
The factor  $\hat T^{\text{tw4}}_{3g\to p\bar{p}} $ has the following schematic  structure 
\begin{equation}
\hat T^{\text{tw4}}_{3g\to p\bar{p}}(x_{2})\simeq \sum_{i}\frac{1}{x_{i}}%
X_{i}(x_{i})+X_{i}(x_{i})\frac{\partial }{\partial k_{\bot i}},  \label{Ptw4}
\end{equation}%
where $X_{i}$ denote one of twist-4 DAs in Eq.(\ref{asXi}). The powers $%
1/x_{i}$ originate from the inverse derivatives $(in\partial )^{-1}$ in the
twist-4 operator in Eq.(\ref{Otw4}). From Eqs.(\ref{asXi}) and (\ref{Ptw4}%
) can be also seen that the terms with transverse derivatives are always
suppressed by factor  $v^{2}$ comparing to  terms with $1/x_i$ and therefore can be neglected. 
Then one finds that $\hat T^{\text{tw4}}_{3g\to p\bar{p}}(x_{2})\sim \mathcal{O}(v^{0})$  which gives 
\begin{equation}
\hat T_{3g\to p\bar{p}}\sim y_{1}y_{2}\bar{y}_{2}~\hat T^{\text{tw4}}_{3g\to p\bar{p}}(x_{2})\sim \mathcal{O}(v^{2}).  \label{Psoft}
\end{equation}

Consider now the sum of the heavy quark subdiagrams $D(k_{i}^{\prime },k_{j})$%
. Performing expansions with respect to small fractions $x_{1}$ and $y_{1}$
one obtains that the most singular terms appear from the diagrams describing
the attachments of the ultrasoft gluon to external vertices on the heavy quark
line. It is convenient to divide such diagrams into two groups: the soft
gluon vertex is associated with the external heavy quark or or with the exteranl heavy
antiquark. Then the sum of all relevant diagrams reads 
\begin{align}
\frac14\text{Tr}\left[(1-\Dsl{\omega})\Dsl{\epsilon}_\psi D^{\mu_1\mu_2\mu_3}\right]  =
\frac{\frac14\text{Tr}\left[ (1-\NEG{\omega})\NEG%
{\epsilon}_{\psi }\gamma ^{\mu_1}(m_Q\NEG{\omega}+m_Q-\Dsl{k}_{1}-\Dsl{k}_{1}^{\prime})
D_{h}^{\mu _{2}\mu _{3}}\right] }
{\left[ -P\left( k_{1}+k_{1}^{\prime}\right) +2(k_{1}k_{1}^{\prime })\right] } 
\notag \\
+\frac{\frac14\text{Tr}\left[ (1-\NEG{\omega})\NEG{\epsilon}_{\psi }
D_{h}^{\mu_{2}\mu _{3}}(-m_Q\NEG{\omega}+m_Q+\Dsl{k}_{1}+\Dsl{k}_{1}^{\prime })\gamma ^{\mu_{1}}\right]}
{\left[ -P\left( k_{1}+k_{1}^{\prime }\right)+2(k_{1}k_{1}^{\prime })\right] },
\end{align}
where$~D_{h}^{\mu _{2}\mu _{3}}$ describes the sum of the subdiagrams with
the hard gluons. The expansion with respect to the small fractions yields 
\begin{align}
&\frac14\text{Tr}\left[(1-\Dsl{\omega})\Dsl{\epsilon}_\psi D^{\mu_1\mu_2\mu_3}\right]_{us}\simeq -\frac{1}{(kk^{\prime })}\frac{1}{\left(
x_{1}+y_{1}\right) }\frac14\text{Tr}\left[ (1-\NEG{\omega})\NEG{\epsilon}_{\psi
}\gamma ^{1}(m\NEG{\omega}+m)D_{h}^{\mu _{2}\mu _{3}}\right]
\notag \\
&\mskip 100mu -\frac{1}{(kk^{\prime })}\frac{1}{\left( x_{1}+y_{1}\right) }\frac14\text{Tr}\left[
(1-\NEG{\omega})\NEG{\epsilon}_{\psi }D_{h}^{\mu _{2}\mu _{3}}(-m\NEG%
{\omega}+m)\gamma ^{\mu _{1}}\right] +\mathcal{O}(v^{0})
\\[4mm]
&\mskip 100mu \simeq -\frac{1}{(kk^{\prime })}\frac{2m\omega ^{1}}{\left(
x_{1}+y_{1}\right) }\text{Tr}\left[ (1-\NEG{\omega})\NEG{\epsilon}_{\psi
}D_{h}^{\mu _{2}\mu _{3}}\right] 
\notag \\
& \mskip 200mu-\frac{1}{(kk^{\prime })}\frac{-2m\omega ^{1}}{\left( x_{1}+y_{1}\right) }%
\text{Tr}\left[ (1-\NEG{\omega})\NEG{\epsilon}_{\psi }D_{h}^{\mu _{2}\mu
_{3}}\right] +\mathcal{O}(v^{0})=\mathcal{O}(v^{0}).  \label{Dsoft}
\end{align}
We see that each separate term  in $D_{us}$ has the contribution of order $v^{-2}$ due to
the factor $1/(x_{1}+y_{1})$ but these terms cancel in the sum. 
Substituting (\ref{Psoft}) and (\ref{Dsoft}) in (\ref{Jsoft}) one obtains
\begin{equation}
J_{us}\sim \int_{0}^{\eta }\frac{dx_{1}}{x_{1}}\int_{0}^{\eta }dy_{1}\times 
\mathcal{O}(v^{0})\sim v^{2},
\end{equation}
Hence  the contribution of the ultrasoft region is power suppressed. 

The same conclusion  is also true for other regions where $x_{i}\sim y_{i}\sim v^{2}$.  This result  
is in agreement with the Coulomb limit described by the potential NRQCD \cite{Pineda:1997bj, Pineda:1997ie, Beneke:1997zp, Brambilla:1999qa, Brambilla:1999xf, Brambilla:2004jw}.  In this case  the ultrasoft gluon vertices
are  suppressed by the small velocity $v$.  

\section{ The convolution integrals }
\lab{AppC}

Here we present  the results for the convolution integrals. All the relevant moments $\varphi_{ij}$, $\lambda_{1}$ and $\eta_{10,11}$  are multiplicatively renormalisable  and their scale dependence are not shown for simplicity. 
After integrations over the momentum fractions the following expressions have been obtained
\bea
I_{0}&=&120^2 \{ 0.1054+0.2282\varphi_{11}+0.4846\varphi _{20}-0.0165\varphi _{22} +0.3958\varphi _{10}^{2}+0.6273\varphi_{10}\varphi_{21} 
\nonumber \\
&& +0.370~\varphi _{11}^{2} +0.5413\varphi _{11}\varphi _{20} -0.0147\varphi _{11}\varphi _{22}+0.6371\varphi _{20}^2-0.0456\varphi _{20}\varphi _{22}
\nonumber \\
&&+0.2283\varphi _{21}^{2} -0.0004\varphi _{22}^{2} \},
\eea
\bea
 J_{1}[V_{1},\mathcal{V}_{i}] + J_{3}[V_{1},\mathcal{A}_{i}] =120\{ P_0[\varphi_{ij}]+ \lambda _{1}/ f_{N }  P_1[\varphi_{ij},\eta_{1k}] \},
\eea
where
\bea
&& P_0[\varphi_{ij}]= 23.6524
-31.2198 \varphi_{10} 
-16.9824 \varphi_{11}
+91.7657 \varphi_ {20}
-25.9159 \varphi_{21}
-9.1100\varphi_{22}
 \nonumber \\ &&
+89.8059\varphi_{10}^2
 -64.5699 \varphi_{10}  \varphi_{11}
 -24.9722 \varphi_{10} \varphi_{20}
+75.5738 \varphi_{10} \varphi_{21}
-32.4481  \varphi_{10}  \varphi_{22}
\nonumber \\ &&
+11.5448  \varphi_{11}^2
-96.2095 \varphi_{11}  \varphi_{20}
+87.5753 \varphi_{11}  \varphi_{21}
+9.6300 \varphi_{11} \varphi_{22}
+61.6547 \varphi_{20}^2
\nonumber \\ &&
 -7.3150 \varphi_{20}  \varphi_{21}
  -21.9330\varphi_{20}  \varphi_{22}
+12.6407 \varphi_{21}^2
 +11.2552 \varphi_{21}  \varphi_{22}
+ 0.1585 \varphi_{22}^2,
\\[2mm]
&&P_1[\varphi_{ij},\eta_{1k}]=
 57.5592  \varphi_{10}
 -19.1864\varphi_{11}
 -2.45327\varphi_{20}
+36.799  \varphi_{21}
-4.9061\varphi_{22}
\nonumber  \\&&
+\eta_{11} (-41.3263
+31.0521\varphi_{10}
-10.3507 \varphi_{11}
-91.8392 \varphi_{20}
+23.7966  \varphi_{21}
+9.72038\varphi_{22})
 \nonumber \\&& 
+\eta_{10} (
 -149.845\varphi_{10}
+49.9484  \varphi_{11}
+1.2259 \varphi_{20}
-18.3879 \varphi_{21}
+ 2.4517 \varphi_{22}).
\eea
\begin{align}
J_{2}[A_{1},\mathcal{V}_{i}]+J_{4}[A_{1},\mathcal{A}_{i}]=120\{(\varphi_{10}+\varphi_{11}) \tilde P_0[\varphi_{ij},\eta_{1k}]+(\varphi_{20}+5\varphi_{21}+2\varphi_{22}) \tilde P_1[\varphi_{ij},\eta_{1k}]\},
\end{align}%
where 
\bea
\tilde P_0[\varphi_{ij},\eta_{1k}]&=&
31.2198
+29.9353 \varphi_{10}
+34.6345 \varphi_{11}
+4.6600\varphi_{20}
-46.0060 \varphi_{21}
-8.1761  \varphi_{22}
\nonumber \\
&+&\lambda_1/f_N (
19.1864 
 -49.9484\eta_{10}
 -31.0521\eta_{11}
),
\\[2mm]
\tilde P_1[\varphi_{ij},\eta_{1k}]&=&
6.0598 
+7.5620 \varphi_{10}
-14.9913\varphi_{11}
+4.7182 \varphi_{20}
+1.6796 \varphi_{21}
-1.1247 \varphi_{22}
\nonumber \\
&+&\lambda_1/f_N (
2.4533
-1.2259 \eta_{10}
-4.7593\eta_{11}).
\eea
\begin{align}
J_{5}[T_{1},\mathcal{T}_{ij}]=120\{ P_{t0}[\varphi_{ij}]+ \lambda_1 \eta_{11} /f_N P_{t1}[\varphi_{ij}] \},
\label{J5num}
\end{align}
where 
\bea
P_{t0}[\varphi_{ij}]&=&47.3049 +28.4748 \varphi_{11}+195.6510\varphi_{20}+8.7661 \varphi_{21}+6.0192\varphi_{22}
\nonumber \\
&+&92.3587\varphi_{11}^2 -213.08156\varphi_{11} \varphi_{20}-66.7746 \varphi_{11} \varphi_{21}-57.0375\varphi_{11} \varphi_{22}
\nonumber \\
&+&132.7459\varphi_{20}^2+26.8001\varphi_{20} \varphi_{21}-27.2425 \varphi_{20} \varphi_{22}-0.2418 \varphi_{21}\varphi_{22}
\nonumber \\
&-&4.1821\varphi_{22}^2,
\\[2mm]
P_{t1}[\varphi_{ij}]&=&-82.6526-82.8057 \varphi_{11}-193.1971\varphi_{20}+0.4035 \varphi_{22}.
\eea

\section*{Acknowledgements}
I am grateful to I.~Anikin,  A.~Manashov, V.~Braun, G. Duplan\v{c}i\'c  for  useful discussions   and A.~Kivel for the help with the text of the paper.


\begin{thebibliography}{99}                                                                                               


\bibitem{Brambilla:2004wf}
  N.~Brambilla {\it et al.} [Quarkonium Working Group],
  hep-ph/0412158.
  
\bibitem{Brambilla:2010cs}
  N.~Brambilla {\it et al.},
  Eur.\ Phys.\ J.\ C {\bf 71} (2011) 1534
  doi:10.1140/epjc/s10052-010-1534-9
  [arXiv:1010.5827 [hep-ph]].
  
\bibitem{Bodwin:1994jh} G.~T.~Bodwin, E.~Braaten and G.~P.~Lepage, 
Phys.\ Rev.\ D \textbf{51} (1995) 1125 [Phys.\ Rev.\ D \textbf{55} (1997)
5853] [hep-ph/9407339]. 

\bibitem{Lepage:1992tx} G.~P.~Lepage, L.~Magnea, C.~Nakhleh, U.~Magnea and
K.~Hornbostel, 
Phys.\ Rev.\ D \textbf{46} (1992) 4052 [hep-lat/9205007]. 
  
  
\bibitem{Brodsky:1981kj}
  S.~J.~Brodsky and G.~P.~Lepage,
  Phys.\ Rev.\ D {\bf 24} (1981) 2848.
  doi:10.1103/PhysRevD.24.2848


\bibitem{Chernyak:1983ej}
  V.~L.~Chernyak and A.~R.~Zhitnitsky,
  Phys. Rept.  {\bf 112} (1984) 173.
  doi:10.1016/0370-1573(84)90126-1
  
  
\bibitem{Tanabashi:2018oca}
  M.~Tanabashi {\it et al.} [Particle Data Group],
  Phys.\ Rev.\ D {\bf 98} (2018) no.3,  030001.
  doi:10.1103/PhysRevD.98.030001
  
\bibitem{Claudson:1981fj}
  M.~Claudson, S.~L.~Glashow and M.~B.~Wise,
  Phys.\ Rev.\ D {\bf 25} (1982) 1345.
 doi:10.1103/PhysRevD.25.1345
  


\bibitem{Peruzzi:1977pb}
  I.~Peruzzi {\it et al.},
  Phys.\ Rev.\ D {\bf 17} (1978) 2901.
  doi:10.1103/PhysRevD.17.2901
  
\bibitem{Brandelik:1979hy}
  R.~Brandelik {\it et al.} [DASP Collaboration],
  Z.\ Phys.\ C {\bf 1} (1979) 233.
  doi:10.1007/BF01440224
  
\bibitem{Eaton:1983kb}
  M.~W.~Eaton {\it et al.},
  Phys.\ Rev.\ D {\bf 29} (1984) 804.
  doi:10.1103/PhysRevD.29.804
  
\bibitem{Pallin:1987py}
  D.~Pallin {\it et al.} [DM2 Collaboration],
  Nucl.\ Phys.\ B {\bf 292} (1987) 653.
  doi:10.1016/0550-3213(87)90663-8
  
\bibitem{Bai:2004jg}
  J.~Z.~Bai {\it et al.} [BES Collaboration],
  Phys.\ Lett.\ B {\bf 591} (2004) 42
  doi:10.1016/j.physletb.2004.04.022
  [hep-ex/0402034].
  
\bibitem{Ambrogiani:2004uj}
  M.~Ambrogiani {\it et al.} [Fermilab E835 Collaboration],
  Phys.\ Lett.\ B {\bf 610} (2005) 177
  doi:10.1016/j.physletb.2005.01.093
  [hep-ex/0412007].
  
\bibitem{Ablikim:2006aw}
  M.~Ablikim {\it et al.} [BES Collaboration],
  Phys.\ Lett.\ B {\bf 648} (2007) 149
  doi:10.1016/j.physletb.2007.02.029
  [hep-ex/0610079].
  
\bibitem{Ablikim:2012eu}
  M.~Ablikim {\it et al.} [BESIII Collaboration],
  Phys.\ Rev.\ D {\bf 86} (2012) 032014
  doi:10.1103/PhysRevD.86.032014
  [arXiv:1205.1036 [hep-ex]].
  
  
\bibitem{Carimalo:1985mw}
  C.~Carimalo,
  Int.\ J.\ Mod.\ Phys.\ A {\bf 2} (1987) 249.
  doi:10.1142/S0217751X87000107
  
\bibitem{Murgia:1994dh}
  F.~Murgia and M.~Melis,
  Phys.\ Rev.\ D {\bf 51} (1995) 3487
  doi:10.1103/PhysRevD.51.3487
  [hep-ph/9412205].
  
\bibitem{Ping:2002uj}
  R.~G.~Ping, H.~C.~Chiang and B.~S.~Zou,
  Phys.\ Rev.\ D {\bf 66} (2002) 054020.
  doi:10.1103/PhysRevD.66.054020
 
\bibitem{Braun:2000kw}
  V.~Braun, R.~J.~Fries, N.~Mahnke and E.~Stein,
  Nucl.\ Phys.\ B {\bf 589} (2000) 381
   Erratum: [Nucl.\ Phys.\ B {\bf 607} (2001) 433]
  doi:10.1016/S0550-3213(00)00516-2, 10.1016/S0550-3213(01)00254-1
  [hep-ph/0007279].
  
\bibitem{Braun:2006hz}
  V.~M.~Braun, A.~Lenz and M.~Wittmann,
  Phys.\ Rev.\ D {\bf 73} (2006) 094019
  doi:10.1103/PhysRevD.73.094019
  [hep-ph/0604050].
  
\bibitem{Anikin:2013aka}
  I.~V.~Anikin, V.~M.~Braun and N.~Offen,
  Phys.\ Rev.\ D {\bf 88} (2013) 114021
  doi:10.1103/PhysRevD.88.114021
  [arXiv:1310.1375 [hep-ph]].
  
 
  
\bibitem{Bali:2015ykx}
  G.~S.~Bali {\it et al.},
  JHEP {\bf 1602} (2016) 070
  doi:10.1007/JHEP02(2016)070
  [arXiv:1512.02050 [hep-lat]].
  
\bibitem{Bali:2019ecy}
  G.~S.~Bali {\it et al.} [RQCD Collaboration],
  Eur.\ Phys.\ J.\ A {\bf 55} (2019) no.7,  116
  doi:10.1140/epja/i2019-12803-6
  [arXiv:1903.12590 [hep-lat]].
  
  
\bibitem{Chernyak:1984bm}
  V.~L.~Chernyak and I.~R.~Zhitnitsky,
  Nucl.\ Phys.\ B {\bf 246} (1984) 52.
  doi:10.1016/0550-3213(84)90114-7
  
\bibitem{Chernyak:1987nv}
  V.~L.~Chernyak, A.~A.~Ogloblin and I.~R.~Zhitnitsky,
  Z.\ Phys.\ C {\bf 42} (1989) 583
   [Yad.\ Fiz.\  {\bf 48} (1988) 1398]
   [Sov.\ J.\ Nucl.\ Phys.\  {\bf 48} (1988) 889].
  doi:10.1007/BF01557664
   
     
\bibitem{Belitsky:2002kj}
  A.~V.~Belitsky, X.~d.~Ji and F.~Yuan,
  Phys.\ Rev.\ Lett.\  {\bf 91} (2003) 092003
  doi:10.1103/PhysRevLett.91.092003
  [hep-ph/0212351].

  
  
\bibitem{Braun:2008ia}
  V.~M.~Braun, A.~N.~Manashov and J.~Rohrwild,
  Nucl.\ Phys.\ B {\bf 807} (2009) 89
  doi:10.1016/j.nuclphysb.2008.08.012
  [arXiv:0806.2531 [hep-ph]].

    
\bibitem{Anikin:2013yoa}
  I.~V.~Anikin and A.~N.~Manashov,
  Phys.\ Rev.\ D {\bf 89} (2014) no.1,  014011
  doi:10.1103/PhysRevD.89.014011
  [arXiv:1311.3584 [hep-ph]].
 
 
\bibitem{Bolz:1996sw}
  J.~Bolz and P.~Kroll,
  Z.\ Phys.\ A {\bf 356} (1996) 327
  doi:10.1007/s002180050186
  [hep-ph/9603289].

\bibitem{King:1986wi}
  I.~D.~King and C.~T.~Sachrajda,
  Nucl.\ Phys.\ B {\bf 279} (1987) 785.
  doi:10.1016/0550-3213(87)90019-8
  
\bibitem{Stefanis:1992nw}
  N.~G.~Stefanis and M.~Bergmann,
  Phys.\ Rev.\ D {\bf 47} (1993) R3685
  doi:10.1103/PhysRevD.47.R3685
  [hep-ph/9211250].
 
 
\bibitem{Eichten:1995ch}
  E.~J.~Eichten and C.~Quigg,
  Phys.\ Rev.\ D {\bf 52} (1995) 1726
  [hep-ph/9503356].
  


\bibitem{Pineda:1997bj} A.~Pineda and J.~Soto, 
Nucl.\ Phys.\ Proc.\ Suppl.\ \textbf{64} (1998) 428 [hep-ph/9707481]. 

\bibitem{Pineda:1997ie} A.~Pineda and J.~Soto, 
Phys.\ Lett.\ B \textbf{420} (1998) 391 [hep-ph/9711292]. 


\bibitem{Beneke:1997zp} M.~Beneke and V.~A.~Smirnov, 
Nucl.\ Phys.\ B \textbf{522} (1998) 321 [hep-ph/9711391]. 


\bibitem{Brambilla:1999qa} N.~Brambilla, A.~Pineda, J.~Soto and A.~Vairo, 
Phys.\ Rev.\ D \textbf{60} (1999) 091502 [hep-ph/9903355]. 

\bibitem{Brambilla:1999xf} N.~Brambilla, A.~Pineda, J.~Soto and A.~Vairo, 
Nucl.\ Phys.\ B \textbf{566} (2000) 275 [hep-ph/9907240]. 

\bibitem{Brambilla:2004jw} N.~Brambilla, A.~Pineda, J.~Soto and A.~Vairo, 
Rev.\ Mod.\ Phys.\ \textbf{77} (2005) 1423 [hep-ph/0410047]. 


\end{thebibliography}
\end{document}